\documentclass[%
 reprint,
amsmath,amssymb,
aps,
prc,
showpacs,
longbibliography]{revtex4-1}

\usepackage{graphicx}
\usepackage{epsfig}
\usepackage{dcolumn}
\usepackage{bm}

\begin{document}

\title{Two- and three-cluster decays of light nuclei with the hyperspherical harmonics}

\author{V. S. Vasilevsky}
\email{vsvasilevsky@gmail.com}
\author{Yu. A. Lashko}
\email{ylashko@gmail.com}
\author{G. F. Filippov}

\address{Bogolyubov Institute for Theoretical Physics, \\14-b Metrolohichna str., 03143, Kiev, Ukraine}

\begin{abstract}
We consider a set of three-cluster systems ($^{4}$He, $^{7}$Li, $^{7}$Be, $^{8}$Be, $^{10}$Be) within  a microscopic model  which
involves the hyperspherical harmonics to represent intercluster motion. We
selected such three-cluster systems which have at least one binary channel. Our
aim is to study whether the hyperspherical harmonics are able and under what
conditions to describe two-body channel(s) (nondemocratic motion) or they are
suitable for describing three-cluster continuum only (democratic motion). It
is demonstrated that a rather restricted number of the hyperspherical
harmonics allows us to describe bound states and scattering states in two-body continuum for a three-cluster system.

\end{abstract}


\maketitle

\section{Introduction}

The hyperspherical harmonics (HH) method is a powerful tool for solving
many-body problems in different branches of quantum physics, namely atomic,
molecular and nuclear physics. In the orthodox realization of the method, a
many-body problem is reduced to a finite or an infinite set of coupled channel
problems, representing the many-body Schr\"{o}dinger equation as a set of
coupled one-dimensional differential equations. Efficiency of the method has
been repeatedly demonstrated by numerous investigations of few-body problems.
Besides, this method has been constantly advanced by creating a more reliable
and universal technique for description of the discrete and continuous spectra
of many-body systems.

One of the direction for the HH method development is to use a full set of
oscillator functions which are labelled by quantum numbers of the
hyperspherical harmonics method. We will call them hyperspherical oscillator functions.

In the present paper, we study different channels of decay of three-cluster
systems \ and ability of the hyperspherical oscillator functions to describe
democratic and nondemocratic decay channels. In literature (see, for instance,
Refs. \cite{1980NCimA..59..283A, 1989NuPhA.505..215B,
2007PhRvC..75e1304C}) a democratic decay is a synonym for three-body
decay or full disintegration of a three-body system. This type of the decay is
also called a "true" \cite{1980CzJPh..30.1090J} or "truly"
\cite{1971RSPTA.270..197G} three--body scattering. Contrary to the
\ democratic decay, a nondemocratic decay stands for a decay of compound system
into two fragments provided that one of the fragments is represented by a
bound state of a two-body subsystem. In what follows we will consider only the
dominant three-body configurations of light atomic nuclei.

Note that an oscillator basis is a conventional set of functions which are
involved in many nuclear models, such as a traditional many-body shell model,
the resonating group method, novel ab initio No-Core Shell Model and many others.

One of the advantages of the basis of hyperspherical oscillator functions that it
allows to circumvent the problems enforced by the Pauli principle. The basis
simplifies numerical solving of coupled channel differential equations by
reducing them to an algebraic matrix form.

Let us consider the following nuclei and appropriate (dominant) three-cluster
configurations (3CC), as well as binary decay channels (2CC) (see
Table \ref{Tab:NuclConfig}). In Table \ref{Tab:NuclConfig} we indicated only
those two-cluster decay channels of the three-cluster systems which have a bound state in the corresponding
two-cluster subsystem. For instance, for $^{7}$Li we take into account
channels $\alpha+t$ and $^{6}$Li$+n$, and we omit channel $^{5}$He$+d$ because
there are no bound states in $^{5}$He subsystem. In other words, we disregard
those binary channels whose threshold energy exceeds the three-cluster
threshold. In Table \ref{Tab:NuclConfig} we also made references to our papers
where the corresponding nucleus has been investigated.%

\begin{table}[htbp] \centering
\begin{ruledtabular}
\caption{List of nuclei, their dominant three-cluster configurations and
dominant two-body decay channels. }%
\begin{tabular}
[c]{cccccc}
Nucleus & 3CC & 2CC 1 & 2CC 2 & 2CC 3 & Paper\\\hline
$^{4}$He & $d+p+n$ & $^{3}$H$+p$ & $^{3}$He$+n$ & $d+d$ &  \\ 
$^{7}$Li & $\alpha +d+n$ & $\alpha +t$ & $^{6}$Li$+n$ &  & \cite%
{2009PAN....72.1450N} \\ 
$^{7}$Be & $\alpha +d+p$ & $\alpha +^{3}$He & $^{6}$Li$+p$ &  & \cite%
{2009NuPhA.824...37V} \\ 
$^{8}$Be & $\alpha +^{3}$H$+p$ & $^{7}$Li$+p$ & $\alpha +\alpha $ &  &  \\
$^{10}$Be & $\alpha +\alpha +^{2}n$ & $^{6}$He$+\alpha $ & $^{8}$Be$+^{2}n$
&  & \cite{Lashko201778} \\ 
\end{tabular}
\label{Tab:NuclConfig}%
\end{ruledtabular}
\end{table}%

We are going to study the eigenspectrum of a microscopic hamiltonian of 
the above-mentioned three-cluster systems. For this
aim we will construct matrix elements of the hamiltonian between many-particle 
 oscillator functions. Diagonalization of the matrix yields eigenvalues
and the corresponding eigenfunctions. Some of the obtained eigenvalues
represent bound states of the compound system, however the largest part of the
eigenvalues are discretized states in two- or three-cluster continuum.
The number of the eigenvalues and their density in the energy range in
question depend on the number of oscillator functions involved in calculation
and naturally on the properties of nucleus under consideration.

\begin{figure}[ptbh]
\begin{center}
\includegraphics[width=\columnwidth]%
{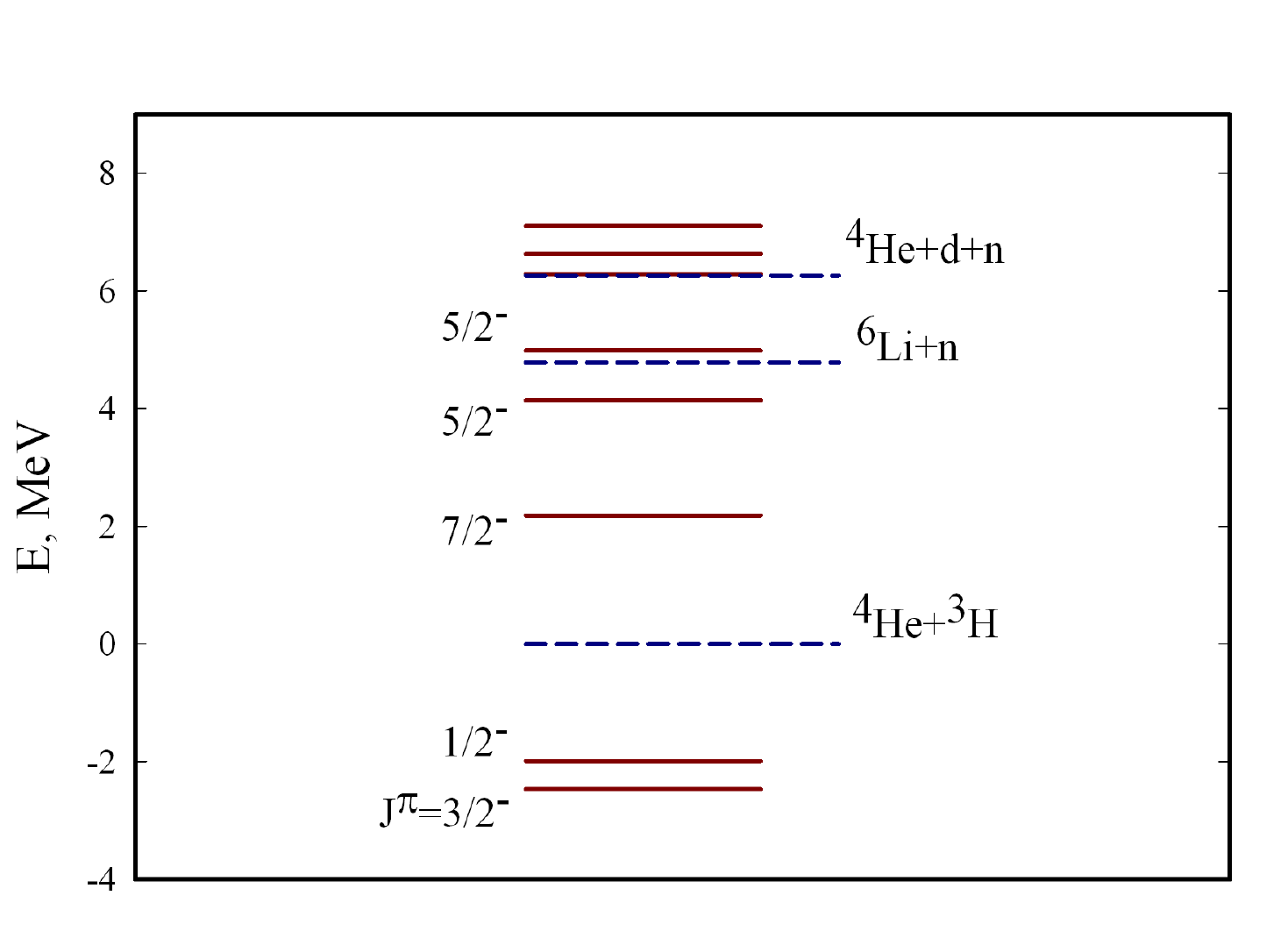}%
\caption{Experimental spectrum of $^{7} $Li. }%
\label{Fig:Spectr7LiExp}%
\end{center}
\end{figure}

To formulate more clearly our aim, let us consider experimental spectrum of
$^{7}$Li (see Ref. \cite{2002NuPhA.708....3T}), one of the nuclei we plan to investigate. In Fig.
\ref{Fig:Spectr7LiExp} we display not only the well-known bound and resonance
states in $^{7}$Li, but also the energy of the \ lowest two-body decay
thresholds ($^{4}$He$+^{3}$H and $^{6}$Li$+n$) and one three-body decay
threshold ($\alpha+d+n$). Within the present paper we will study if the
hyperspherical harmonics are able to reproduce the bound states of $^{7}$Li,
which have dominant two-cluster structure $^{4}$He$+^{3}$H, and how many
hyperspherical harmonic have to be involved in calculations to achieve this
goal. It is also very interesting to examine whether hyperspherical harmonics
allow one to study continuous spectrum between $^{4}$He$+^{3}$H and $^{6}%
$Li$+n$ threshold, and also between two-body $^{6}$Li$+n$\ and three-body
$\alpha+d+n$ thresholds. If so, then what is the required number of
hyperspherical harmonics to solve this problem.%

The problem, we are going to tackle in this paper, have been addressed in
other way in \cite{PhysRevC.26.2288, PhysRevC.26.2301,
2001PhR...347..373N, 2015FBS....56..753D} within the so-called
the hyperspherical adiabatic approximation. This approximation works perfectly
for a system of three structureless particles and allows one to determine
correctly effective potentials for three-body system and the energies of two-
and three-body decay thresholds. However, this approximation is not
suitable for three interacting clusters, as the Pauli principle generates
nonlocal and energy dependent potentials.

The method, which we are going to employ, was formulated in Ref.
\cite{2001PhRvC..63c4606V}. It was successfully applied to study of bound and
resonance states of the Borromean nuclei (such as $^{6}$He, $^{9}%
$Be, $^{12}$C) and resonance states of nuclei with prominent
three-cluster features (such as $^{5}$H, $^{6}$Be, $^{9}%
$B). All these nuclei were considered as three-cluster systems. The
present paper may be considered as a step forward in creating of an unified
microscopic model for describing both binary and three-cluster channels.

In this paper we will study wave functions of the deeply- and weakly-bound states
accounted both from two- and three-cluster threshold. As for the weakly-bound states,
we will demonstrate that  a binary structure is revealed in the three-cluster wave
function of the pseudo-bound states whose energy is close to the corresponding binary
decay threshold. Despite the fact that to obtain convergent results for the energies
of resonance states lying between the lowest binary decay threshold and three-cluster
decay threshold of a three-cluster system one needs a large basis of hyperspherical
harmonics, it is possible to see the evidence of two-cluster structure in the three-cluster
wave function of a pseudo-bound state yet with a rather restricted set of hyperspherical harmonics.

This article is organized as follows. In Section \ref{Sec:Method} we present a
brief review of a microscopic three-cluster model which exploits the
hyperspherical harmonics. Main attention in this section is paid to the
asymptotic form of three-cluster wave functions describing two- and
three-cluster decay of a compound system. In Section \ref{Sec:Results}, we
analyze convergence of spectrum of three-cluster systems and study
peculiarities of wave functions of bound and pseudo-bound states in different
asymptotic regimes. Brief conclusions are presented in Section
\ref{Sec:Conclusions}.

\section{Method \label{Sec:Method}}

First of all we need to introduce coordinates which determine relative
position of clusters in coordinate space. The most suitable variables we
believe are the Jacobi vectors $\mathbf{x}$ and $\mathbf{y}$. One can
introduce three different sets (or different trees) of Jacobi vectors. Within
the Faddeev formalism for three particles or within coupled channel formalism
for three clusters, one has to use all three sets of Jacobi vectors. In what
follows we stick to one tree of the Jacobi vectors. We will discuss latter why
only one set of the Jacobi vectors is sufficient for our purposes in the
hyperspherical harmonics formalism.

We start model formulation with an explicit form of wave function for a
system consisting of three $s$-clusters%
\begin{equation}
\Psi_{E,LM_{L}}=\widehat{\mathcal{A}}\left\{  \Phi_{1}\left(  A_{1}\right)
\Phi_{2}\left(  A_{2}\right)  \Phi_{3}\left(  A_{3}\right)  \psi_{E,LM_{L}%
}\left(  \mathbf{x},\mathbf{y}\right)  \right\}.  \label{eq:001}%
\end{equation}
This is a traditional form of a wave function of the resonating group method
for systems, when at least one cluster consists of two and more nucleons. The
internal structure of clusters ($\alpha$= 1, 2, 3) is described by the
antisymmetric and translationally invariant wave functions $\Phi_{\alpha
}\left(  A_{\alpha}\right)  $. Function $\Phi_{\alpha}\left(  A_{\alpha
}\right)  $ is a wave function of the many-particle shell model with the most
compact configuration of nucleons. These functions are selected in such way to
provide in the most economical manner a fairly good description of \ the main
internal properties (bound state energy, cluster size) of each cluster. The
antisymmetrization operator $\widehat{\mathcal{A}}$ makes antisymmetric the
wave function of the compound three-cluster system, which is of paramount
importance for the energy region under consideration. Since all functions
$\Phi_{\alpha}\left(  A_{\alpha}\right)  $ are fixed, to calculate a spectrum
and wave functions of the compound system one has to determine a wave function
of intercluster motion $\psi_{LM_{L}}\left(  \mathbf{x},\mathbf{y}\right)  $.
This function is also translationally invariant and depends on two Jacobi
vectors $\mathbf{x}$ and $\mathbf{y}$, locating relative position of clusters
in the space. By using angular orbital momentum reduction, we represent this
function as an infinite series
\begin{equation}
\psi_{E,LM_{L}}\left(  \mathbf{x},\mathbf{y}\right)  \Rightarrow\sum_{\lambda
,l}\psi_{E;\lambda,l;L}\left(  x,y\right)  \left\{  Y_{\lambda}\left(
\widehat{\mathbf{x}}\right)  Y_{l}\left(  \widehat{\mathbf{y}}\right)
\right\}  _{LM_{L}}, \label{eq:001B}%
\end{equation}
where $\widehat{\mathbf{x}}$ and $\widehat{\mathbf{y}}$ are unit vectors, and
$\lambda$ and $l$ are the partial angular momenta associated with vectors
$\mathbf{x}$ and $\mathbf{y}$ respectively.

Note that the form of Eq. (\ref{eq:001}) implies that the total orbital
momentum $L$ is a good quantum number. It also means that in what follows we
disregard the noncentral components of nucleon-nucleon interaction (for the
sake of simplicity).

\subsection{Hyperspherical coordinates and basis}

Wave functions of intercluster motion $\psi_{\lambda,l;L}\left(  x,y\right)  $
obey an infinite set of the two dimension (in terms of variables $x\ $and $y$)
integro-differential equations. To solve this equation we make use of the
hyperspherical coordinates and hyperspherical harmonics. There are several
schemes for introducing hyperspherical coordinates or more exactly \ -
hyperspherical angles $\Omega_{5}$, as hyperspherical radius is determined
uniquely. Each choices of the hyperspherical angles invokes a certain set of
five quantum numbers numerating hyperspherical harmonics. Different sets of
the hyperspherical angles, however, have three common quantum numbers: the
hypermomentum $K$, the total orbital momentum $L$ and its projection $M_{L}$
on $z$ axis.

We make use of the hyperspherical harmonics in the form suggested by Zernike
and Brinkman in Ref. \cite{kn:ZB}, because this form is very simple, it does not involve bulky
calculations and the
quantum numbers \ have clear physical meaning. To introduce the Zernike - Brinkman hyperspherical harmonics, we
need to determine the hyperspherical coordinates. Instead of six variables
$\mathbf{x}$ and $\mathbf{y}$ or $x$, $y$, and two unit vectors $\widehat
{\mathbf{x}}$ and $\widehat{\mathbf{y}}$ we introduce a hyperspherical
radius
\begin{equation}
\rho=\sqrt{x^{2}+y^{2}} \label{eq:004}%
\end{equation}
and a hyperspherical angle%
\[
\theta=\arctan\left(  \frac{x}{y}\right).
\]
At a given value of $\rho$, this angle determines relative length of the
vectors $\mathbf{x}$ and $\mathbf{y}$%
\begin{equation}
x=\rho \cos\theta,\quad y=\rho \sin\theta,\quad\theta\in\left[  0,\pi/2\right]  .
\label{eq:005}%
\end{equation}
For small values of $\theta$, the length of vector $y$ is close to zero, and
vector $x$ has maximal value $x\approx\rho$. When the hyperspherical angle
$\theta$ is close to $\pi/2$,  the length of vector $x$ is very small and
vector $y$ is at its maximum.

The represented set of the hyperspherical angles is very popular scheme of the
hyperspherical coordinates for investigating three-body \cite{kn:Zhuk93E,
2001FBS....30...39V, 2001FBS....31...13C},  and three-cluster
systems \cite{2001PhRvC..63c4606V, 2001PhRvC..63c4607V,
2010PPN....41..716N, 2004NuPhA.740..249K}.

In new coordinates%
\begin{equation}
\psi_{E,LM_{L}}\Rightarrow\sum_c\phi_{E,c}\left(\rho\right)  \mathcal{Y}_{c}\left( \Omega_{5}\right), 
\label{eq:006}
\end{equation}
where $\mathcal{Y}_{c}\left(  \Omega_{5}\right)  $ stands\ for the product
\begin{equation}
\mathcal{Y}_{c}\left(  \Omega_{5}\right)  =\chi_{K}^{\left(  \lambda,l\right)
}\left(  \theta\right)  \left\{  Y_{\lambda}\left(  \widehat{\mathbf{x}%
}\right)  Y_{l}\left(  \widehat{\mathbf{y}}\right)  \right\}  _{LM_{L}}
\label{eq:007}%
\end{equation}
and represents a hyperspherical harmonic for a three-cluster channel
\begin{equation}
c=\left\{
K,\lambda,l,L\right\}.
\label{eq:007A}%
\end{equation}
The hyperspherical harmonic $Y_{c}\left(  \Omega
_{5}\right)  $ is a function of five angular variables $\theta,\theta_{1}%
,\phi_{1},\theta_{2},\phi_{2}$. Definition of all
components of the hyperspherical harmonic $Y_{c}\left(  \Omega_{5}\right)  $
can be found, for instance, in Ref. \cite{2001PhRvC..63c4606V}. Being a
complete basis, the hyperspherical harmonics account for any shape of the
three-cluster triangle and its orientation. Thus they account for 
all possible modes of relative motion of \ three interacting clusters.

As for the hyperradial wave functions $\phi_{E,c}\left(  \rho\right)  $, they
obey a system of differential equations with local effective potentials for
three structureless particles, \ or a set of integro-differential equations
with nonlocal effective potentials for three clusters. The latter can be
represented as%
\begin{widetext}
\begin{equation}
\sum_{\widetilde{c}}\left[  \delta_{c,\widetilde{c}}\widehat{T}_{K}\phi
_{E,c}\left(  \rho\right)  +\int d\widetilde{\rho}\widetilde{\rho}%
^{5}V_{c,\widetilde{c}}\left(  \rho,\widetilde{\rho}\right)  \phi
_{E,\widetilde{c}}\left(  \widetilde{\rho}\right)  \right]  =E\sum
_{\widetilde{c}}\int d\widetilde{\rho}\widetilde{\rho}^{5}\mathcal{N}_{c,\widetilde{c}%
}\left(  \rho,\widetilde{\rho}\right)  \phi_{E,\widetilde{c}}\left(
\widetilde{\rho}\right),  \label{eq:008}%
\end{equation}
\end{widetext}
where $\mathcal{N}_{c,\widetilde{c}}\left(  \rho,\widetilde{\rho}\right)$ is a norm kernel and
\begin{equation}
\widehat{T}_{K}=-\frac{\hbar^{2}}{2m}\left[  \frac{\partial^{2}}{\partial
\rho^{2}}+\frac{5}{\rho}\frac{\partial}{\partial\rho}-\frac{K\left(
K+4\right)  }{\rho^{2}}\right].  \label{eq:008A}%
\end{equation}
As one can see, the Pauli principle leads to appearance of the energy
dependent part in the effective potential (the right-hand side of equations
(\ref{eq:008})). To simplify solving a set of equations (\ref{eq:008}), we
invoke a full set of oscillator functions to expand the sought wave
function
\[
\psi_{E,LM_{L}}=\sum_{n_{\rho},c}C_{E;n_{\rho},c}\left\vert n_{\rho}%
,c\right\rangle.
\]
This reduces a set of integro-differential equations (\ref{eq:008}) to an
algebraic form, i.e. to the system of linear algebraic equations%
\begin{equation}
\sum_{\widetilde{n}_{\rho},\widetilde{c}}\left[  \left\langle n_{\rho
},c\left\vert \mathcal{H}_{c,\widetilde{c}}  \right\vert \widetilde{n}_{\rho},\widetilde
{c}\right\rangle -E\left\langle n_{\rho},c\left\vert \mathcal{N}_{c,\widetilde{c}}
 \right\vert\widetilde{n}_{\rho},\widetilde
{c}\right\rangle \right]  C_{E;\widetilde{n}_{\rho},\widetilde{c}} = 0,
\label{eq:009}%
\end{equation}
where $\mathcal{H}_{c,\widetilde{c}}$  is a hamiltonian kernel%
\[
\mathcal{H}_{c,\widetilde{c}}=\mathcal{H}_{c,\widetilde{c}}\left(
\rho,\widetilde{\rho}\right)  =\delta_{c,\widetilde{c}}\widehat{T}_{K}%
\delta\left(  \rho-\widetilde{\rho}\right)  +V_{c,\widetilde{c}}\left(
\rho,\widetilde{\rho}\right).
\]

Oscillator functions for three-cluster configuration are determined as
\begin{eqnarray}
\left\vert n_{\rho},c\right\rangle =\left\vert n_{\rho},K;\lambda
,l;L\right\rangle  \label{eq:010}=R_{n_{\rho}%
K}\left(  \rho,b\right)  \mathcal{Y}_{c}\left(  \Omega_{5}\right),
\end{eqnarray}
where $R_{n_{\rho},K}\left(  \rho,b\right)  $ is an oscillator function%
\begin{eqnarray}
&  & R_{n_{\rho},K}\left(  \rho,b\right)   \nonumber \\
 &  =& \left(  -1\right)  ^{n_{\rho}%
}\mathcal{N}_{n_{\rho},K}r^{K}\exp\left\{  -\frac{1}{2}r^{2}\right\}
L_{n_{\rho}}^{K+3}\left(  r^{2}\right)  ,\label{eq:010A}\\
r  &  =& \rho/b,\quad\mathcal{N}_{n_{\rho},K}=b^{-3}\sqrt{\frac{2\Gamma\left(
n_{\rho}+1\right)  }{\Gamma\left(  n_{\rho}+K+3\right)  }},\nonumber
\end{eqnarray}
and $b$ is an oscillator length.

System of equations (\ref{eq:009}) can be solved numerically by imposing
restrictions on the number of hyperradial excitations $n_{\rho}$ and on the
number of hyperspherical channels $c_{1}$, $c_{2}$, \ldots, $c_{N_{ch}}$. The
diagonalization procedure may be used to determine energies and wave functions
of the bound states. However, the proper boundary conditions have to be implemented
to calculate elements of the scattering $S$-matrix and corresponding
functions of continuous spectrum. \ Boundary conditions or asymptotic behaviour
of wave functions for democratic and nondemocratic decay of a compound
three-cluster system will be considered in Section \ref{Sec:Asympt}.

Wave function (\ref{eq:010}) belongs to the oscillator shell with the number
of oscillator quanta $N_{os}=2n_{\rho}+K$. It is convenient to numerate the
oscillator shells by $N_{sh}$ ( = 0, 1, 2, \ldots), which we determine as%
\[
N_{os}=2n_{\rho}+K=2N_{sh}+K_{\min},%
\]
where $K_{\min}=L$ for normal parity states $\pi=\left(  -1\right)  ^{L}$ and
$K_{\min}=L+1$ for abnormal parity states $\pi=\left(  -1\right)  ^{L+1}$.
Thus we account oscillator shells starting from a "vacuum" shell ($N_{sh}=0$)
with minimal value of the hypermomentum $K_{\min}$ compatible with a given
total orbital momentum $L$.

It is worthwhile noticing that Eq. (\ref{eq:009}) contains the norm kernel matrix
$\left\Vert \left\langle n_{\rho},c\left\vert \mathcal{N}_{c,\widetilde{c}} \right\vert\widetilde{n}_{\rho},\widetilde
{c}\right\rangle \right\Vert $, which is also called the matrix of the
antisymmetrization operator. Appearance of this matrix in the equation means
that the oscillator basis functions (\ref{eq:010}) are not orthonormal due to
the Pauli principle. Moreover, these functions may be linear dependent which
leads to appearance of the Pauli forbidden states. This problem has been
numeriously addressed in the past, see, for example, Refs.
\cite{1997PAN....60..343V}, \cite{VNCh2001E}.

Existence of the Pauli forbidden states requires that  the system of equations
(\ref{eq:009}) to be solved in two steps. In the first step, one needs to
diagonalize the norm kernel matrix $\left\Vert \left\langle n_{\rho},c\left\vert  \mathcal{N}_{c,\widetilde{c}} \right\vert\widetilde{n}_{\rho
},\widetilde{c}\right\rangle \right\Vert $. This matrix has a block structure.
Matrix $\left\Vert \left\langle n_{\rho},c\left\vert  \mathcal{N}_{c,\widetilde{c}} \right\vert\widetilde{n}_{\rho}%
,\widetilde{c}\right\rangle \right\Vert $ has a very large number of zero matrix
elements. Nonvanishing matrix elements of this matrix are
 overlaps of basis functions of the same oscillator shell. This statement
can be expressed as the following condition:%
\[
2n_{\rho}+K=2\widetilde{n}_{\rho}+\widetilde{K}%
\]
Such block structure of the overlap matrix significantly simplifies
calculations of its eigenstates $\Lambda_{N_{sh},\alpha}$ ($\alpha$ = 1, 2,
\ldots) and corresponding eigenfunctions $O_{n_{\rho},c}^{N_{sh},\alpha}$.
These eigenfunctions form an orthogonal matrix which transform the original
basis functions $\left\vert n_{\rho},c\right\rangle $ to a new set of
functions%
\[
\left\vert N_{sh},\alpha\right\rangle =\sum_{n_{\rho},c\in N_{sh}}O_{n_{\rho
},c}^{N_{sh},\alpha}\left\vert n_{\rho},c\right\rangle .
\]
We check whether a particular eigenvalue $\Lambda_{N_{sh},\alpha}$ equals to
zero. If so, this eigenvalue is the Pauli forbidden state and has to be
eliminated from our Hilbert space. Otherwise, this eigenstate belongs to the
part of the total three-cluster Hilbert space spanned by the Pauli allowed
states. The first stage of solving a set of Eqs. (\ref{eq:009}) is completed by constructing the normalized Pauli
allowed states $\left\vert N_{sh},c_{a}\right\rangle =$ $\left\vert
N_{sh},\alpha\right\rangle /\sqrt{\Lambda_{N_{sh},\alpha}}$, where index
$c_{a}$ numerates the Pauli allowed states on the oscillator shell $N_{sh}$.

In the second step, one needs to transform the matrix of hamiltonian from original
basis of functions $\left\vert n_{\rho},c\right\rangle $ to the basis of the
normalized Pauli allowed states $\left\vert N_{sh},c_{a}\right\rangle $.

For numerical solution of Eq. (\ref{eq:009}) one has to construct matrix
elements of a microscopic hamiltonian with a selected nucleon-nucleon
potential and the norm kernel matrix. We do not dwell on this problem since the
appropriate method of constructing such matrices was formulated in Ref.
\cite{2001PhRvC..63c4606V}.

\subsection{Strategy of investigation \label{Sec:Strateg}}

To achieve the goals formulated above we
  carry out our investigations in the following steps:

\begin{itemize}
\item We calculate matrix elements of a hamiltonian between three-cluster
oscillator functions

\item By solving an eigenvalue problem, we obtain spectrum of bound and
pseudo-bound states, and the corresponding wave functions in discrete representation.

\item To study the nature of the obtained solutions, we construct correlation
functions and calculate the weight of different oscillator shells in wave
functions of bound and pseudo-bound states

\item Finally, we analyse the structure of the wave functions in discrete and
coordinate representations and study their asymptotic behaviour.
\end{itemize}

Some remarks should be made to explain and justify this strategy. First, such way of investigating  a three-cluster system allows us to avoid the application of the necessary boundary conditions which leads to the very bulky and time-consuming calculations. Diagonalization procedure enables us in a rather simple way to obtain wave functions of bound and scattering states in the internal region where effects of cluster-cluster interaction are very strong. By increasing the number of oscillator functions we obtain more correct description of the bound states; gradually approaching to the exact value of the energy and to the correct shape of a wave function, mainly improving its asymptotic tail.

Meanwhile, we have got a different situation for continuous spectrum states. The extension of oscillator basis allows one to scan the continuous spectrum and to study the internal part of wave functions for different energies of the spectrum. The internal part of wave functions represents the exact wave function with a specific ``boundary condition''. This will be latter discussed in more detail.   By concluding we note, that the digonalization procedure of a huge but restricted hamiltonian matrix is very often used to study spectra of light nuclei with the \textit{ab initio} no-core shell model.

\subsection{Asymptotic behaviour of the wave functions \label{Sec:Asympt}}

Let us consider an asymptotic behaviour of wave function (\ref{eq:001}). It is
more appropriate to introduce an analog of the Faddeev amplitudes $\psi
_{L}^{\left(  \alpha\right)  }\left(  \mathbf{x}_{\alpha},\mathbf{y}_{\alpha
}\right)  $
\begin{eqnarray}
& & \Psi_{E,LM_{L}} \label{eq:001A}  \\
 &  = & \sum_{\alpha=1}^{3}\widehat{\mathcal{A}}\left\{  \Phi
_{1}\left(  A_{1}\right)  \Phi_{2}\left(  A_{2}\right)  \Phi_{3}\left(
A_{3}\right)  \psi_{E,LM_{L}}^{\left(  \alpha\right)  }\left(  \mathbf{x}%
_{\alpha},\mathbf{y}_{\alpha}\right)  \right\} \nonumber \\
&  = & \sum_{\alpha=1}^{3}\sum_{\lambda_{\alpha},l_{\alpha}}\widehat{\mathcal{A}%
}\left\{  \Phi_{1}\left(  A_{1}\right)  \Phi_{2}\left(  A_{2}\right)  \Phi
_{3}\left(  A_{3}\right)  \right.  \nonumber\\
&  \times & \left.  \psi_{E,\lambda_{\alpha},l_{\alpha};L}^{\left(
\alpha\right)  }\left(  x_{\alpha},y_{\alpha}\right)  \left\{  Y_{\lambda
_{\alpha}}\left(  \widehat{\mathbf{x}}_{\alpha}\right)  Y_{l_{\alpha}}\left(
\widehat{\mathbf{y}}_{\alpha}\right)  \right\}  _{LM_{L}}\right\}  \nonumber
\end{eqnarray}
to study the asymptotic properties of three-cluster systems. In Eq.
(\ref{eq:001A}), $\mathbf{x}_{\alpha}$ is the Jacobi vector, determining the
distance between $\beta$ and $\gamma$ clusters, while $\mathbf{y}_{\alpha}$ is
a Jacobi vector linking the $\alpha$th cluster with the center of mass of the
$\beta$\ and $\gamma$ clusters:
\begin{eqnarray}
\mathbf{x}_{\alpha } & = & \sqrt{\frac{A_{\beta }A_{\gamma }}{A_{\beta
}+A_{\gamma }}}\left( \mathbf{R}_{\beta }-\mathbf{R}_{\gamma }\right) , 
\label{eq:002a}\\
\mathbf{y}_{\alpha } & = & \sqrt{\frac{A_{\alpha }\left( A_{\beta }+A_{\gamma
}\right) }{A_{\alpha }+A_{\beta }+A_{\gamma }}}\left[ \mathbf{R}_{\alpha }-%
\frac{A_{\beta }\mathbf{R}_{\beta }+A_{\gamma }\mathbf{R}_{\gamma }}{%
A_{\beta }+A_{\gamma }}\right] , \label{eq:002b} \\
\mathbf{R}_{\sigma } & = & \frac{1}{A_{\sigma }}\sum_{i\in A_{\sigma }}\mathbf{r%
}_{i}. \nonumber %
\end{eqnarray}
The indexes $\alpha$, $\beta$\ and $\gamma$\ form a cyclic permutation of 1, 2
and 3.

In general, there are four different asymptotic regions in a
three-cluster system. The asymptotic properties of three-body functions in
these regions are thoroughly discussed in Refs. \cite{Merkurev:1976uy,
kn:faddeev+merk93, Friar3partprobl1996}. We follow notations of
Ref. \cite{kn:faddeev+merk93} (see pages 134-136) and denote these asymptotic
regions as $\Omega_{\alpha}$ ($\alpha$= 1, 2, 3) and $\Omega_{0}$. In the
asymptotic regions $\Omega_{\alpha}$ the distance between clusters with
indexes $\beta$\ and $\gamma$ is much smaller than the distance of the third
cluster (with index $\alpha$) to the center of mass of clusters $\beta$\ and
$\gamma$ ($\left\vert \mathbf{x}_{\alpha}\right\vert \ll\left\vert
\mathbf{y}_{\alpha}\right\vert $). This region describes scattering of a
cluster with index $\alpha$ on a bound state of clusters $\beta$\ and $\gamma
$. \ The asymptotic region $\Omega_{0}$ describes situation when all clusters
are well separated, i.e. when intercluster distances $\left\vert
\mathbf{x}_{\alpha}\right\vert \gg a$ ($\alpha$= 1, 2, 3) are larger then the
radius $a$ of a short range interaction.

Consider the asymptotic region $\Omega_{\alpha}$. In this region, the Faddeev
amplitude $\psi_{E,L}^{\left(  \alpha\right)  }\left(  \mathbf{x}_{\alpha
},\mathbf{y}_{\alpha}\right)  $ has the following "two-body" asymptotic form
\begin{eqnarray}
&  & \psi_{E,\lambda_{\alpha},l_{\alpha};L}^{\left(  \alpha\right)  }\left(
x_{\alpha},y_{\alpha}\right) \approx \sum_{\lambda_{\alpha},l_{\alpha}}g_{\lambda_{\alpha},\mathcal{E}%
_{\alpha}}\left(  x_{\alpha}\right)\nonumber\\
& \times &   \left[  \delta_{c_{0},c} \
\psi_{k_{\alpha},l_{\alpha}}^{\left(  -\right)  }\left(  k_{\alpha}y_{\alpha
}\right)  -S_{c_{0},c} \ \psi_{k_{\alpha},l_{\alpha}}^{\left(  +\right)
}\left(  k_{\alpha}y_{\alpha}\right)  \right]  \label{eq:011}%
\end{eqnarray}
for scattering states (when the total energy $E\geq\mathcal{E}_{\alpha}$) and
\begin{eqnarray}
&& \psi_{E,\lambda_{\alpha},l_{\alpha};L}^{\left(  \alpha\right)  }\left(
x_{\alpha},y_{\alpha}\right)   \nonumber \\
&  \approx & \sum_{\lambda_{\alpha},l_{\alpha}%
}A_{c_{0},c}g_{\lambda_{\alpha},\mathcal{E}_{\alpha}}\left(  x_{\alpha
}\right)  W_{-\eta_{\alpha},l_{\alpha}+1/2}\left(  2k_{\alpha}y_{\alpha
}\right) \label{eq:012}\\
&  \approx & \sum_{\lambda_{\alpha},l_{\alpha}}A_{c_{0},c}g_{\lambda_{\alpha
},\mathcal{E}_{\alpha}}\left(  x_{\alpha}\right)  \frac{1}{\left(
\kappa_{\alpha}y_{\alpha}\right)  ^{\eta_{\alpha}+1}}\exp\left\{
-\kappa_{\alpha}y_{\alpha}\right\} \nonumber %
\end{eqnarray}
for bound states (when the total energy $E<\mathcal{E}_{\alpha}$). Here
$\psi_{k_{\alpha},l_{\alpha}}^{\left(  -\right)  }\left(  k_{\alpha}y_{\alpha
}\right)  $ and $\psi_{k_{\alpha},l_{\alpha}}^{\left(  +\right)  }\left(
k_{\alpha}y_{\alpha}\right)  $ are incoming and outgoing waves, respectively,
determined in terms of the well-known regular $F_{l_{\alpha}}\left(
k_{\alpha}y_{\alpha};\eta_{\alpha}\right)  $ and irregular $G_{l_{\alpha}%
}\left(  k_{\alpha}y_{\alpha};\eta_{\alpha}\right)  $ Coulomb functions:%
\begin{widetext}
\begin{eqnarray}
 \psi_{k_{\alpha},l_{\alpha}}^{\left(  \pm\right)  }\left(  k_{\alpha}%
y_{\alpha}\right)    & = &  \left[  G_{l_{\alpha}}\left(  k_{\alpha}y_{\alpha
};\eta_{\alpha}\right)  \pm F_{l_{\alpha}}\left(  k_{\alpha}y_{\alpha}%
;\eta_{\alpha}\right)  \right]  /y_{\alpha}\label{eq:013}\\
&  & \approx_{y_{\alpha}\rightarrow\infty}\exp\left\{  \pm i\left(  k_{\alpha
}y_{\alpha}-l_{\alpha}\frac{\pi}{2}-\eta_{\alpha}\ln\left(  2k_{\alpha
}y_{\alpha}\right)  +\sigma_{l_{\alpha}}\right)  \right\}  /y_{\alpha
},\nonumber
\end{eqnarray}
\end{widetext}
and $W_{-\eta_{\alpha},l_{\alpha}+1/2}\left(  2k_{\alpha}y_{\alpha}\right)  $
is the Whittaker function. In Eqs. (\ref{eq:011}) and (\ref{eq:012}), index
$c=\left\{  \lambda_{\alpha},\mathcal{E}_{\alpha},l_{\alpha}\right\}  $
numerates the current channel, $c_{0}$ indicates the incoming channel,
$\mathcal{E}_{\alpha}$ and $g_{\lambda_{\alpha},\mathcal{E}_{\alpha}}\left(
x_{\alpha}\right)  $ denote energy of the two-cluster bound state and its wave
function, and
\begin{eqnarray}
k_{\alpha}  &  = & \sqrt{\frac{2m\left(  E-\mathcal{E}_{\alpha}\right)  }%
{\hbar^{2}}},\quad\kappa_{\alpha}=\sqrt{\frac{2m\left\vert E-\mathcal{E}%
_{\alpha}\right\vert }{\hbar^{2}}},\label{eq:014A}\\
\quad\eta_{\alpha}  &  = & \frac{Z_{\alpha}\left(  Z_{\beta}+Z_{\gamma}\right)
e^{2}}{\sqrt{\left\vert E-\mathcal{E}_{\alpha}\right\vert }}\sqrt
{\frac{A_{\alpha}\left(  A_{\beta}+A_{\gamma}\right)  }{A}\frac{m}{\hbar^{2}}%
}. \label{eq:014B}%
\end{eqnarray}

Let us turn our attention to the asymptotic region $\Omega_{0}$. We present an
asymptotic form only for neutral clusters (or by neglecting the Coulomb
interaction in asymptotic region). In this case an asymptotic form of
three-cluster wave function is well and unambiguously established. For a bound
state of three-cluster system the wave function $\psi_{E,\lambda_{\alpha
},l_{\alpha};L}^{\left(  \alpha\right)  }\left(  x_{\alpha},y_{\alpha}\right)
$ is%
\begin{eqnarray}
\psi_{E,\lambda_{\alpha},l_{\alpha};L}^{\left(  \alpha\right)  }\left(
x_{\alpha},y_{\alpha}\right)  &=& \psi_{\lambda_{\alpha},l_{\alpha};L}^{\left(
\alpha\right)  }\left(  \rho,\theta_{\alpha}\right)  \label{eq:015A} \\
& \approx & \exp\left\{
-\kappa_{0} \ \rho\right\}  /\rho^{5/2}, \nonumber %
\end{eqnarray}
and for scattering state (full disintegration or breakup)%
\begin{eqnarray}
\psi_{E,\lambda_{\alpha},l_{\alpha};L}^{\left(  \alpha\right)  }\left(
x_{\alpha},y_{\alpha}\right)  &=& \psi_{\lambda_{\alpha},l_{\alpha};L}^{\left(
\alpha\right)  }\left(  \rho,\theta_{\alpha}\right)   \label{eq:015B} \\
&\approx & A_{c_{0}%
,c}\left(  \theta_{\alpha}\right)  \exp\left\{  ik_{0}\ \rho\right\}
/\rho^{5/2}, \nonumber %
\end{eqnarray}
where $A_{c_{0},c}\left(  \theta_{\alpha}\right)  $ is a breakup amplitude
and
\[
k_{0}=\sqrt{\frac{2m\left(  E-\mathcal{E}_{0}\right)  }{\hbar^{2}}}%
,\quad\kappa_{0}=\sqrt{\frac{2m\left\vert E-\mathcal{E}_{0}\right\vert }%
{\hbar^{2}}},
\]
$\mathcal{E}_{0}$ is the three-cluster threshold energy.

As we see the hyperspherical coordinates are involved to express an asymptotic form of wave functions for the democratic decay of three-cluster system or its bound state (see Eqs. (\ref{eq:015A}) and (\ref{eq:015B})). They can be also used to express an
asymptotic behaviour of three-cluster wave functions in all asymptotic regions
by using relations (\ref{eq:005}).

So far, we discussed the asymptotic behaviour of three-cluster wave functions
for different scenario of two- and three-body decays in coordinate
representations. Similar relations can be written in the discrete
representation by using the correspondence between the expansion coefficients and
the wave functions in coordinate space.

Let us assume that we arranged all binary channels in such an order that
$\mathcal{E}_{1}<\mathcal{E}_{2}<\mathcal{E}_{3}$. Besides, we also assume
that all energies are measured from the three-cluster threshold and thus
$\mathcal{E}_{0}=0$. With such definitions, an asymptotic part of bound state
wave functions will be mainly represented by wave function from Eq.
 of the first channel (\ref{eq:012}). Contribution of other binary channels
and three-cluster channel  depends on how far is this state from the
corresponding decay threshold as well. The larger is difference $\left\vert
E-\mathcal{E}_{\sigma}\right\vert $ ($\sigma$ = 0, 1, 2, 3), the smaller is
contribution of the corresponding channel to the asymptotic part of the bound state wave
function. It is obvious that to obtain the deeply bound state (with respect to the
three-cluster threshold), a very small number of the hyperspherical
harmonics is required. For a weakly bound state, the large number of the hyperspherical
harmonics should be involved to reach a necessary precision for the energy of this
state. Wave function of continuous spectrum state with energy $\mathcal{E}%
_{1}<E<\mathcal{E}_{2}$ is expected to be mainly represented by the wave function
of the first channel of the form (\ref{eq:011}). When the energy of this state
approaches energy of the second binary channels, it is natural to expect
a substantial contribution of the second closed channel. More
intrigue situation can be observed for continuous spectrum states when two
binary channels are open, i.e. for the states with energy $\mathcal{E}%
_{2}<E<\mathcal{E}_{3}$. \ Wave function of such state can be
represented either by the first or by the second channels, or by a combination of
both channels. More
complicated situation can be observed for continuous spectrum states when all
binary channels are open ($\mathcal{E}_{3}<E<0$).

\subsection{Oscillator basis}

Note that the first two steps in the strategy mentioned in Section \ref{Sec:Strateg} are common for
many microscopic and semi-microscopic model calculations, since many models
used a square-integrable basis to obtain information about bound and
scattering states. \ Oscillator basis is the most used one among others. Two
main merits of the basis make it popular. First, these basis functions are
orthonormal, they do not create any problem with overfullness or linear
dependence of basis functions. Second, due to the specific properties of
oscillator functions, there is a simple relationship between expansion
coefficients and an original wave function.\ It makes more transparent physical
interpretation of the obtained results.

Now we consider other two advantages of the basis which will be exploited in
what follows. First, oscillator basis allows one to implement correct boundary
conditions in discrete oscillator representation \cite{kn:Fil_Okhr,
kn:Fil81}. Second, diagonalization of  matrix of a microscopic hamiltonian, constructed with the oscillator basis, reveals eigenfunctions with the clear physical properties. To explain the second advantage \ of oscillator basis we
consider a single-channel approximation for two-cluster or two-body system.
Suppose we constructed the matrix elements of  hamiltonian $\widehat{H}$
between oscillator functions $\left\vert n\right\rangle $\ and $\left\vert
m\right\rangle $. By diagonalizing matrix $\left\Vert \left\langle n\left\vert
\widehat{H}\right\vert m\right\rangle \right\Vert $ with dimension of $N\times
N$, we obtain eigenvalues $E_{\nu}$ ($\nu$=0,1, 2, \ldots, $N-1$) and the
corresponding eigenfunctions $\left\{  C_{n}^{\left(  \nu\right)  }\right\}
$. Eigenvalues $E_{\alpha}$ present bound states if $E_{\nu}<0$, and
pseudo-bound states with positive energy $E_{\nu}>0$. It was shown in Refs.
\cite{kn:Heller1, kn:VVS+1983Li7E,2005PPN..36.714F}, that pseudo-bound states are the
states of continuous spectrum states selected from infinite set through the
diagonalization procedure by the condition $C_{N+1}^{\left(  \nu\right)  }=0$.
In other words, these states have the node at a given point. This condition
can be also used to determine phase shifts at these discrete points.
Interesting feature of the pseudo-bound states is that without imposing any
boundary condition one can obtain a set of \textit{discrete} or selected states in
two-cluster continuum. Very important feature of the pseudo-bound states, is
that the wave function of a pseudo-bound state differs by a factor from the
normalized to delta function wave function of continuous spectrum with the
same energy.

\begin{figure}[ptbh]
\begin{center}
\includegraphics[width=\columnwidth]%
{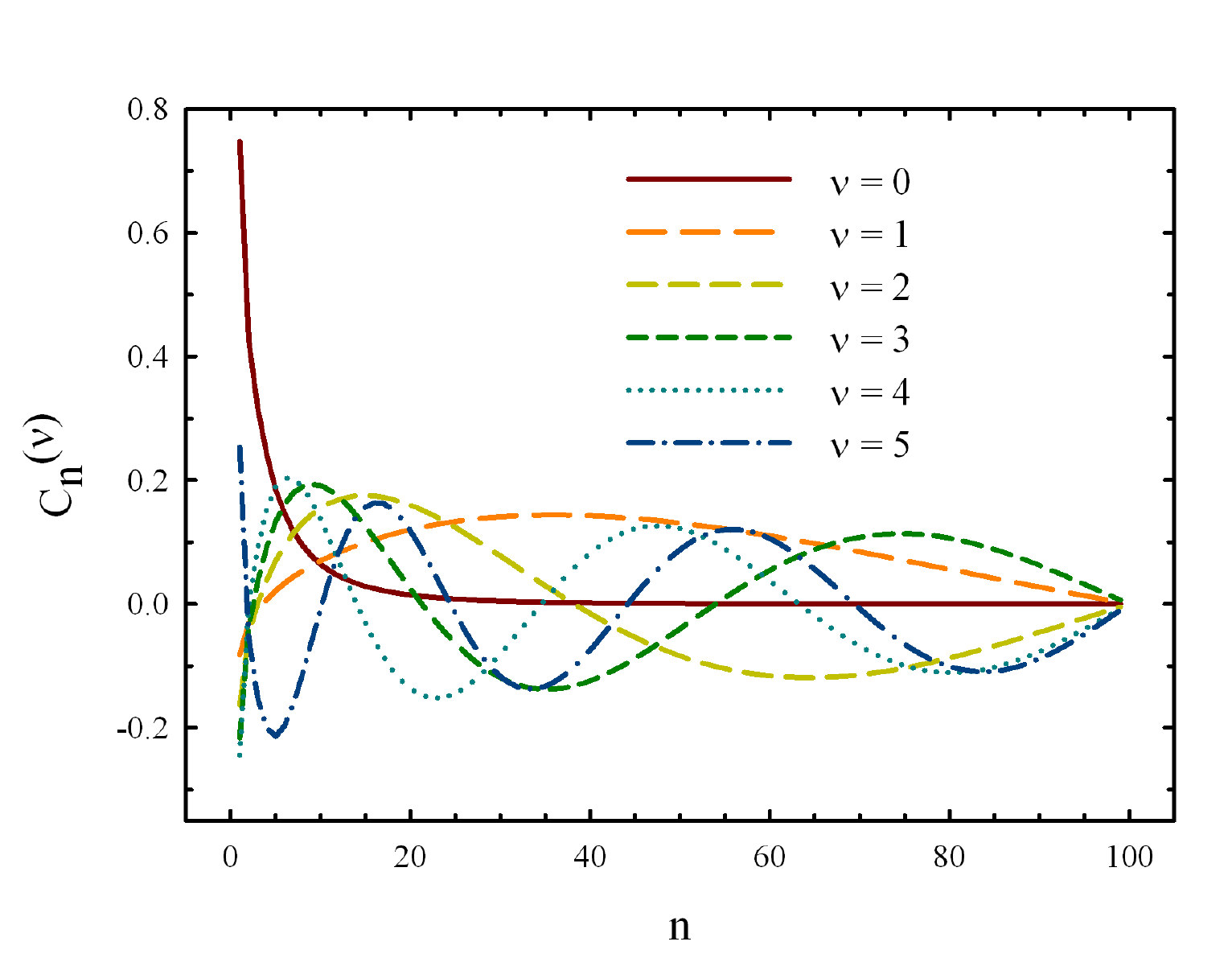}%
\caption{Six eigenfunctions of 3/2$^{-}$ state in $^{7}$Li obtained in
two-cluster $\alpha+t$ approximation.}%
\label{Fig:EigFuns7Li2Cl}%
\end{center}
\end{figure}

To confirm the aforementioned properties of wave functions in the oscillator
representation, we show the following illustration. In Fig.
\ref{Fig:EigFuns7Li2Cl} we display eigenfunctions of two-cluster $\alpha+t$
hamiltonian for the 3/2$^{-}$ state in $^{7}$Li. These functions are
determined with 100 oscillator functions. The first eigenfunction describes
the $^{7}$Li ground state, other five functions represent states of
two-cluster continuum. One can see explicitly that the wave functions of
these states indeed have a node at the end of an interval $n=N+1$. The same is
true for the ground state, however it is not so evident in Fig.
\ref{Fig:EigFuns7Li2Cl}. We do not dwell on the details of such calculations
as they can be found, for instance, in Ref. \cite{Lashko201778}.%

\begin{figure}[ptbh]
\begin{center}
\includegraphics[width=\columnwidth]%
{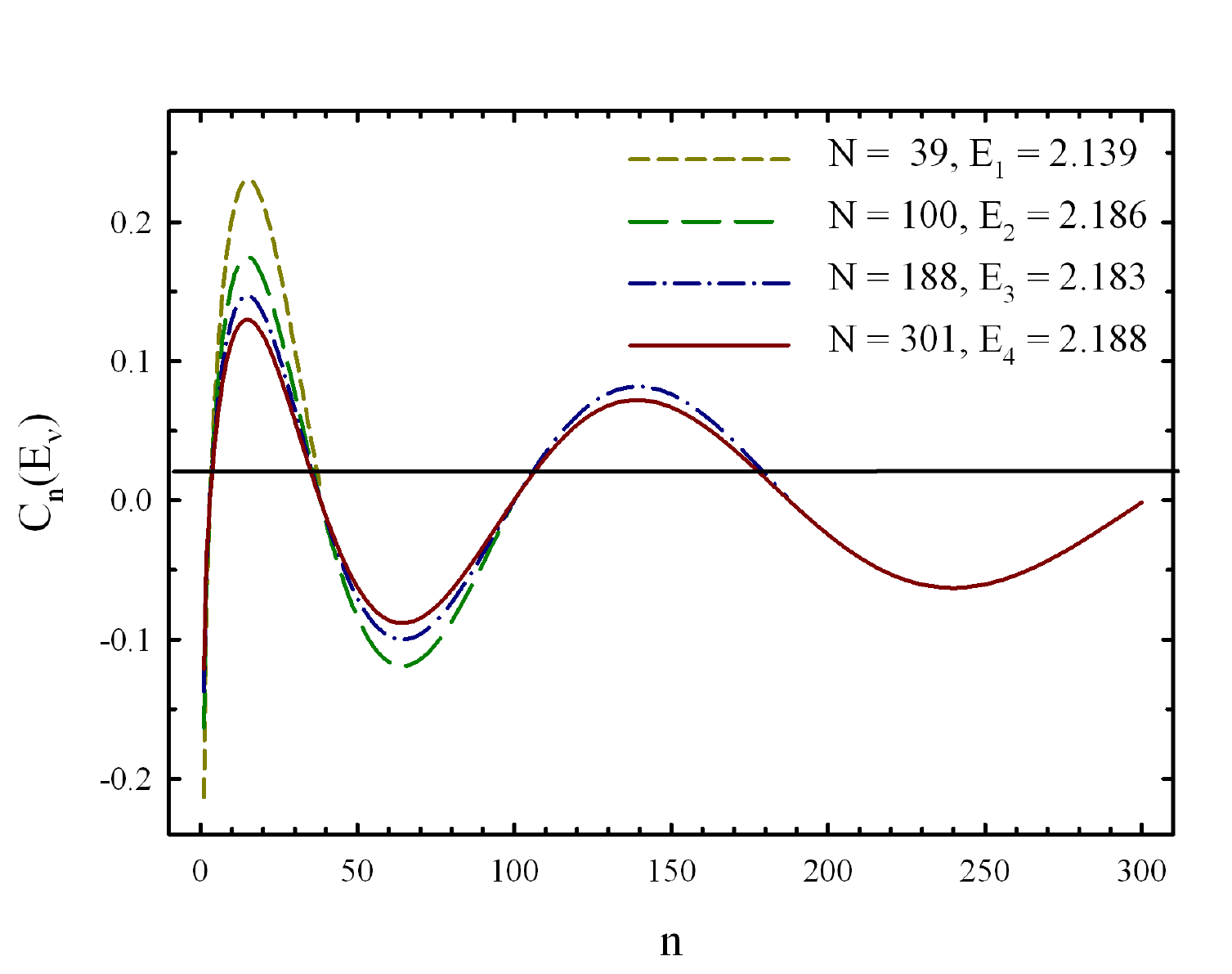}%
\caption{Wave functions of the 3/2$^{-}$ pseudo-bound states in $^{7}$Li$=\alpha+t$ obtained for approximately the same energy and with the different number of oscillator functions.}%
\label{Fig:WaveFuns7LiOB}%
\end{center}
\end{figure}

\begin{figure}[ptbh]
\begin{center}
\includegraphics[width=\columnwidth]%
{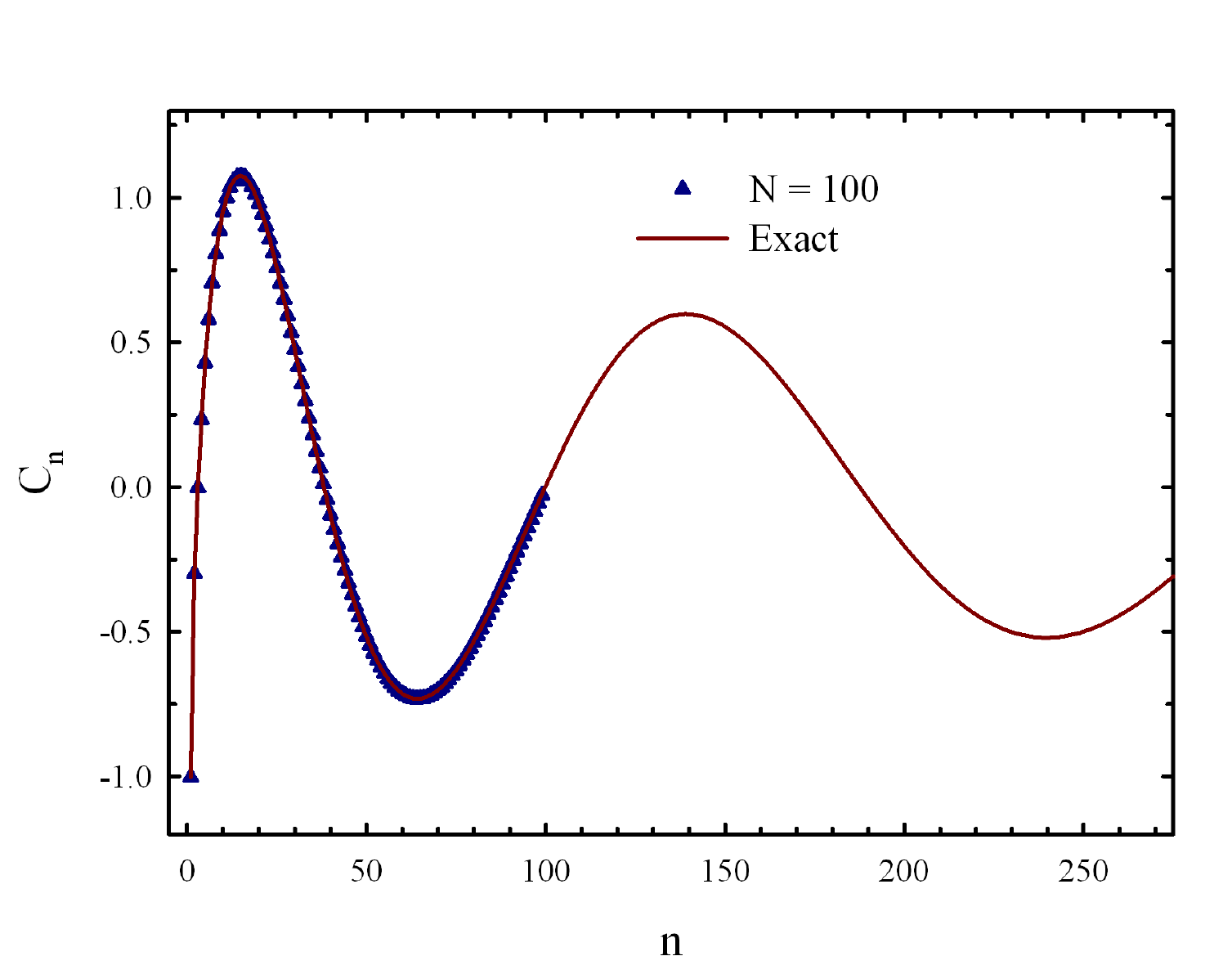}%
\caption{The exact wave function  of the 3/2$^{-}$  state of $^{7}$Li$=\alpha+t$ with energy $E$ = 2.186 MeV compared with the renormalized wave function of the pseudo-bound state, obtained for the same energy and with 100 oscillator functions.}%
\label{Fig:WaveFuns7LiEvsA}%
\end{center}
\end{figure}

To prove shortly that eigenstates of  a two-cluster hamiltonian have clear physical meaning, we consider wave functions of four 3/2$^{-}$ pseudo-bound states in $^{7}$Li. These state are obtained with the different number $N$ of oscillator functions. We selected  and displayed in Fig. \ref{Fig:WaveFuns7LiOB} those pseudo-bound states which have approximately the same energy. Thus one can expect that wave functions of these states have the same structure (shape). Indeed, we see that the selected wave functions have maxima and nodes at the same points of the discrete space. They differ only by the normalization condition
\[
\sum_{n=0}^{N} \left \vert C_n^{\left ( \nu \right )} \right \vert^2 = 1
\]
 and represent the corresponding state in different oscillator spaces $0\le \nu \le N$ ($N$ = 39, 100, 188, 301). In Fig. \ref{Fig:WaveFuns7LiOB}  we indicated the number of the displayed eigenstate $\nu$ and its energy $E_{\nu}$.  If we normalize  four wave functions in the same fashion (for instance, as the wave function with the minimal value of oscillator functions $N$ = 39),  we will obtain four quite close wave functions in the interval $0\le \nu \le 39$, three quite close functions in the interval $0\le \nu \le 100$ and two quite close wave functions in the interval $0\le \nu \le 188$.
Note that these wave functions, obtained with the diagonalization procedure, coincide within a simple normalization factor with the exact wave function of the continuous spectrum state obtained for the same energy  and with the corresponding boundary conditions. This is demonstrated in Fig. \ref{Fig:WaveFuns7LiEvsA}, where we display the exact wave function, obtained for the energy $E$ = 2.186 MeV  by imposing the correct boundary condition, and the renormalized wave function, determined for the same energy with 100 oscillator functions. One can see that both exact and approximate functions coincide in the range $0\le n \le 100$. It is important to note that the exact wave function is determined in the whole oscillator space ($0\le n < \infty$) and normalized by the condition
\[
\sum_{n=0}^{\infty}  C_n^{\left ( E \right ) *} \ C_n^{\left ( \tilde{E} \right )}
= \delta\left (E-\tilde{E} \right ).
\]

To this end, yet another visual confirmation that the diagonalization procedure reveals states with clear physical meaning is presented in Ref. \cite{2005PPN..36.714F}. Fig. 1  (p. 714) of Ref. \cite{2005PPN..36.714F}  demonstrates an approximation of the wave function of the continuum spectrum  by expansion
in the discrete basis with different numbers of basis functions. The above-mentioned figure clearly shows that the larger is the number of basis functions involved, the larger is the region of good approximation of the exact wave function by partial sum. However, the behavior of the wave function in the inner region is well reproduced even with a rather restricted set of basis functions.

Thereby, the wave functions obtained by the diagonalization of the hamiltonian have the same physical meaning as the wave functions obtained with the proper boundary conditions. The only difference is that the former are known only up to a finite point of coordinate space, while the latter are known in the whole region.

We will not dwell on this subject as more details on the J-matrix methods can be found in the book of articles \cite{J-matrix2008B}.


It is also important to note that the methods which use the oscillator basis
are similar to the well-known R-matrix model of nuclear reactions. In these
methods eigenvalues $E_{\nu}$ and eigenfunctions $\left\{  C_{n}^{\left(
\nu\right)  }\right\}  $ are important blocks for construction of the elements
of the scattering matrix and wave functions of single- and many-channel
systems for an arbitrary value of the energy $E$. The explicit formulae relating elements of the $S$-matrix with eigenvalues $E_{\nu}$ and eigenfunctions $\left\{  C_{n}^{\left(
\nu\right)  }\right\}  $ can be found in Refs. \cite{kn:Heller1,1975JMP....16..410Y}. In Refs. \cite{1990JPhG...16.1241M,1998TMP...117.1291Z}, these formulae were extended to the case of many-channel systems described with the hyperspherical basis. Besides, investigations of eigenvalues $E_{\nu}$ and eigenfunctions
$\left\{  C_{n}^{\left(  \nu\right)  }\right\}  $ can be considered as a
simple case of the complex scaling method (see basic definitions of the method
and its recent progress in applications to many-cluster systems in Refs.
\cite{2014PrPNP..79....1M, 2012PThPS.192....1H}), when the rotational
angle equals zero.

\section{Results and discussion \label{Sec:Results}}

We involve the Minnesota potential (MP) \cite{kn:Minn_pot1,
1970NuPhA.158..529R} as a nucleon-nucleon potential in our
calculations. We use a common oscillator length for all clusters. Its value
is selected to minimize the energy of three-cluster threshold. To construct a
wave function of two-cluster relative motion and to determine the energy of a
two-cluster bound state, we employ oscillator basis. Details of two-cluster
calculations can be found in Ref. \cite{Lashko201778}. We make use of 50
oscillator functions to calculate the ground state energies of two-cluster
systems. This number of functions provides correct value of bound state
energies and their parameters (for instance, r.m.s. proton and mass radii and
so on).

In Table \ref{Tab:ParamCalc} we show input parameters of our calculations. It
includes oscillator length $b$ and exchange parameters $u$ of the selected NN potential.%

\begin{table}[tbp] \centering
\begin{ruledtabular}
\caption{Input parameters of calculations for each nucleus.}%
\begin{tabular}
[c]{cccc}
Nucleus & 3CC & $b$, fm & $u$\\\hline
$^{4}$He & $d+\left(  p+n\right)  $ & 1.489 & 0.9600\\
$^{7}$Li & $\alpha+d+n$ & 1.311 & 0.9255\\
$^{7}$Be & $\alpha+d+p$ & 1.311 & 0.9255\\
$^{8}$Be & $\alpha+^{3}H+p$ & 1.311 & 0.9560\\
$^{10}$Be & $\alpha+\alpha+^{2}n$ & 1.356 & 0.9570\\
\end{tabular}
\label{Tab:ParamCalc}%
\end{ruledtabular}
\end{table}%

As in Refs. \cite{2001PhRvC..63f4604V,2007JPhG...34.1955B,2012PhRvC..85c4318V,PhysRevC.96.034322}, 
in the present paper we will vary the hyperspherical momentum in the range 
$L \leq K \leq K_{\max}$, where $K_{\max}$ = 14 for the positive parity states and 
$K_{\max}$ = 13 for the negative parity states. The number of oscillator shells, 
involved in our calculations, is running from zero to $N_{sh,\max}$ = 70. 
These values of $K_{\max}$ and $N_{sh,\max}$, as was shown in Refs. 
\cite{2001PhRvC..63f4604V,2007JPhG...34.1955B,2012PhRvC..85c4318V,PhysRevC.96.034322}, 
provided fairly good description of bound states and resonance states, 
generated by a three-cluster continuum.

\subsection{$^{7}$Li and $^{7}$Be}

Consider evolution of $^{7}$Li spectrum when we involve more and more
hyperspherical harmonics. We consider \ the 3/2$^{-}$ state in both nuclei.
This state is mainly represented by the total orbital momentum $L$ =1. We
restrict ourselves by the only value of the total spin $S$=1/2, and neglect
contribution of negative parity state with total orbital momentum $L$ =2. For
the total orbital momentum $L$ =1, we have only odd values of the
hypermomentum $K$=1, 3, 5\ldots. Thus we represent results with $K$= 1, $K$=3
and so on up to $K_{\max}$=13.

\begin{figure}[ptbh]
\begin{center}
\includegraphics[width=\columnwidth]%
{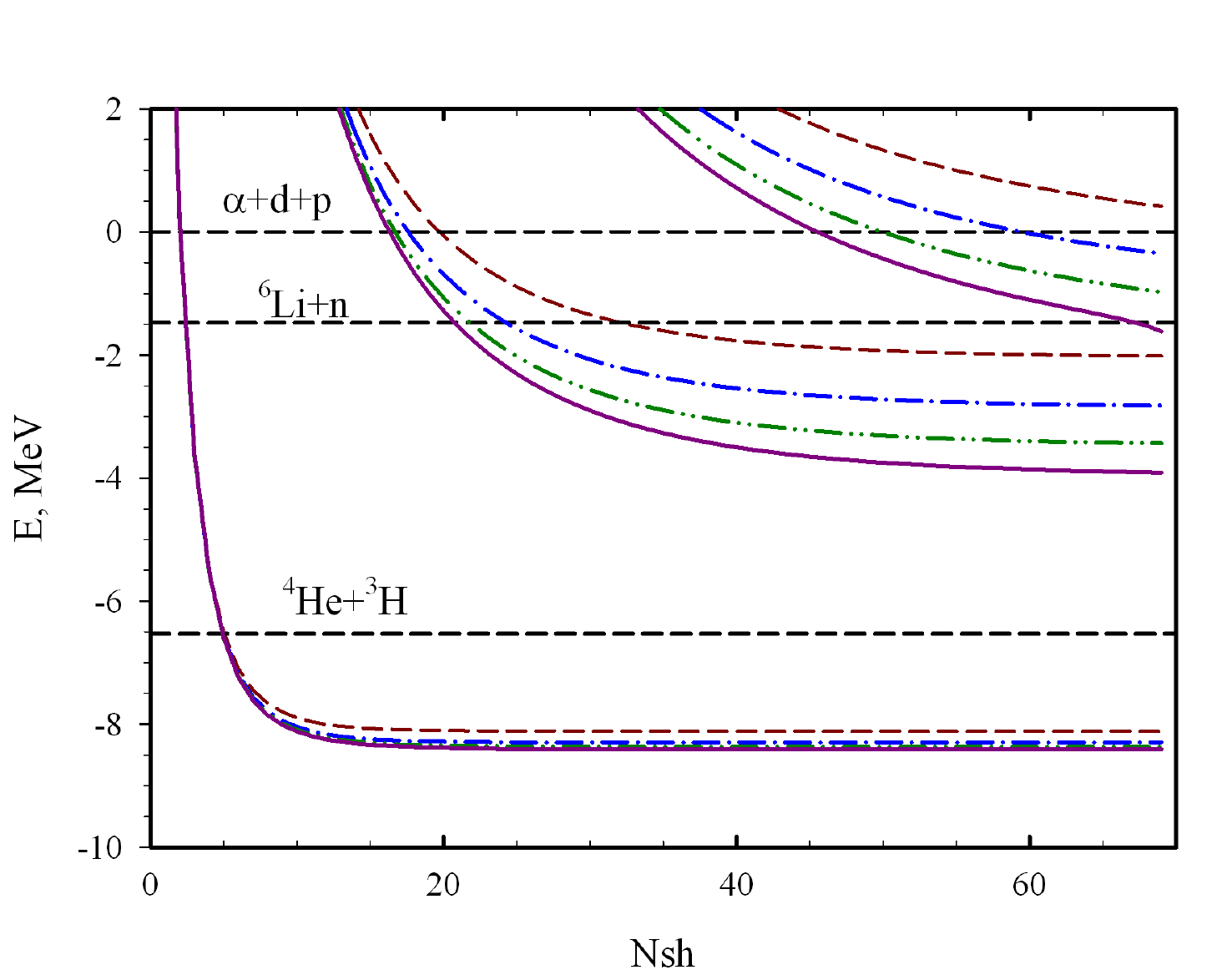}%
\caption{Spectrum of the 3/2$^{-}$ states in $^{7}$Li as a function of
$N_{sh}$ \ and $K_{\max}$. Dashed line - $K_{\max}$= 7, dot-dashed line -
$K_{\max}$= 9, dot-dot-dashed line - $K_{\max}$= 11 and solid line - $K_{\max
}$= 13.}%
\label{Fig:Spectr7Li32MK7_13}%
\end{center}
\end{figure}

Dependence of energy of the 3/2$^{-}$ states in $^{7}$Li on quantum number
$N_{sh}$ is displayed in Fig. \ref{Fig:Spectr7Li32MK7_13}. Here we presented
trajectories of eigenvalues for $^{7}$Li calculated with the hyperspherical
harmonics $K_{\max}$=7, $K_{\max}$=9, $K_{\max}$=11 and $K_{\max}$=13. Fig.
\ref{Fig:Spectr7Li32MK7_13} demonstrates rather fast convergence of the $^{7}%
$Li ground state energy. For the sake of convenience in Fig.
\ref{Fig:Spectr7Li32MK7_13} we connected all discrete points by lines, however
the results are relevant only for discrete values of $N_{sh}$. Thus we need only
restricted number of the hyperspherical harmonics ($K_{\max}$=7) and small
number of hyperradial excitations (or oscillator shells $N_{sh}\leq30$) to
obtain the bound state in $^{7}$Li, i.e. an eigenstate of three-cluster
compound system which lies below the lowest two-cluster threshold $^{4}$He+$^{3}$H.
The first eigenvalue for all values of $K_{\max}$ "scans" two-cluster
continuous spectrum with small values of $N_{sh}$ ($\leq$5) and thus
represents continuous spectrum states in the $^{6}$Li+n channel and in the
$^{4}$He+$^{3}$H channel. The second eigenvalue of the three-cluster
hamiltonian for $7\leq K_{\max}\leq13$ is able to describe continuous spectrum
states in three-cluster continuum and in binary channels continuum. The larger
is the number of the hyperspherical harmonics involved in calculations, the
larger region of two-cluster $^{4}$He+$^{3}$H continuum can be achieved with
these basis functions.%

\begin{figure}[ptbh]
\begin{center}
\includegraphics[width=\columnwidth]{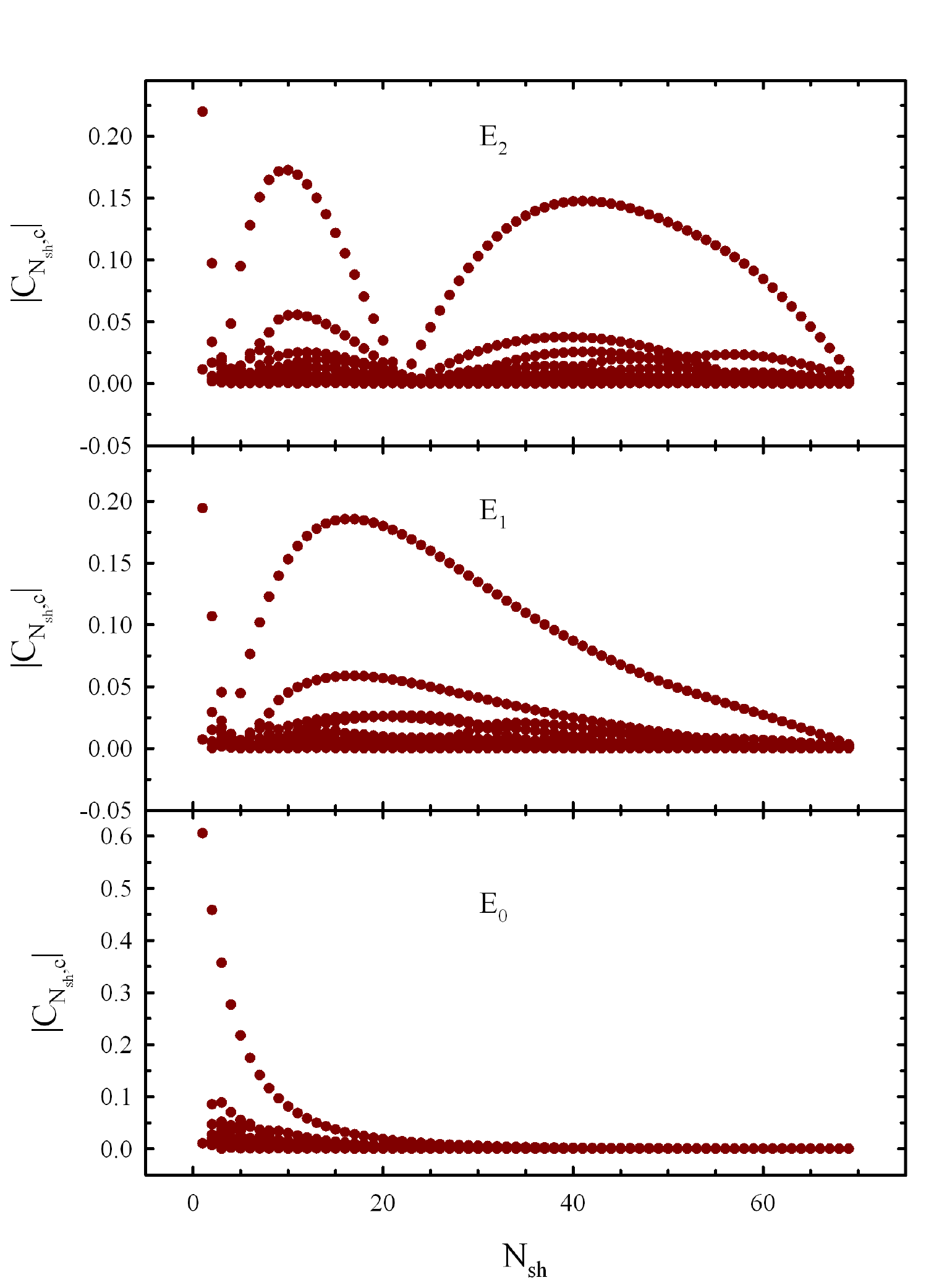}%
\caption{Wave functions of the 3/2$^{-}$ states in $^{7}$Li as a function of
$N_{sh}$.  These function are presented in oscillator representation and calculated with $K_{\max}$ = 13.}
\label{Fig:WaveFuns7LiHHB}%
\end{center}
\end{figure}

It is important to note that the spectrum of $^{7}$Li and other nuclei,
which will be considered bellow, is obtained by solving the generalized
eigenvalue problem represented by Eq. (\ref{eq:009}). By solving this problem we
obtain those states of the compound system  which obey the following
"boundary conditions"%
\[
C_{n_{\rho ,\max }+1,c}=0,
\]%
which are similar to the two-cluster case, considered above. It is
demonstrated in Fig. \ref{Fig:WaveFuns7LiHHB}, where we display three wave functions of the 3/2$^{-}$ states in $^{7}$Li calculated with $K_{\max }$=13. For each value of $N_{sh}\geq 7$, there are 56
channels which are involved in construction of all the functions. As we can see,
the expansion coefficients tend to zero as   $N_{sh}$ approaches
its maximal value. We conjecture some important conclusions from this
figure. First, we involve the large number of the channels $c$, however only a
few of them dominate in the wave functions of bound and pseudo-bound states.
This result is an agreement with the previous results of the model for the wave
functions of three-cluster continuum (see Refs. 
\cite{2001PhRvC..63f4604V,2007JPhG...34.1955B,2012PhRvC..85c4318V,PhysRevC.96.034322}).
 In the latter case, such conclusion was
formulated by analyzing the asymptotic part of wave functions. Second, there
is an apparent similarity between the wave functions of two- and three-cluster
models, which, for example, exhibits in the number of nodes and in the
shape of the wave functions.

\begin{figure}[ptbh]
\begin{center}
\includegraphics[width=\columnwidth]{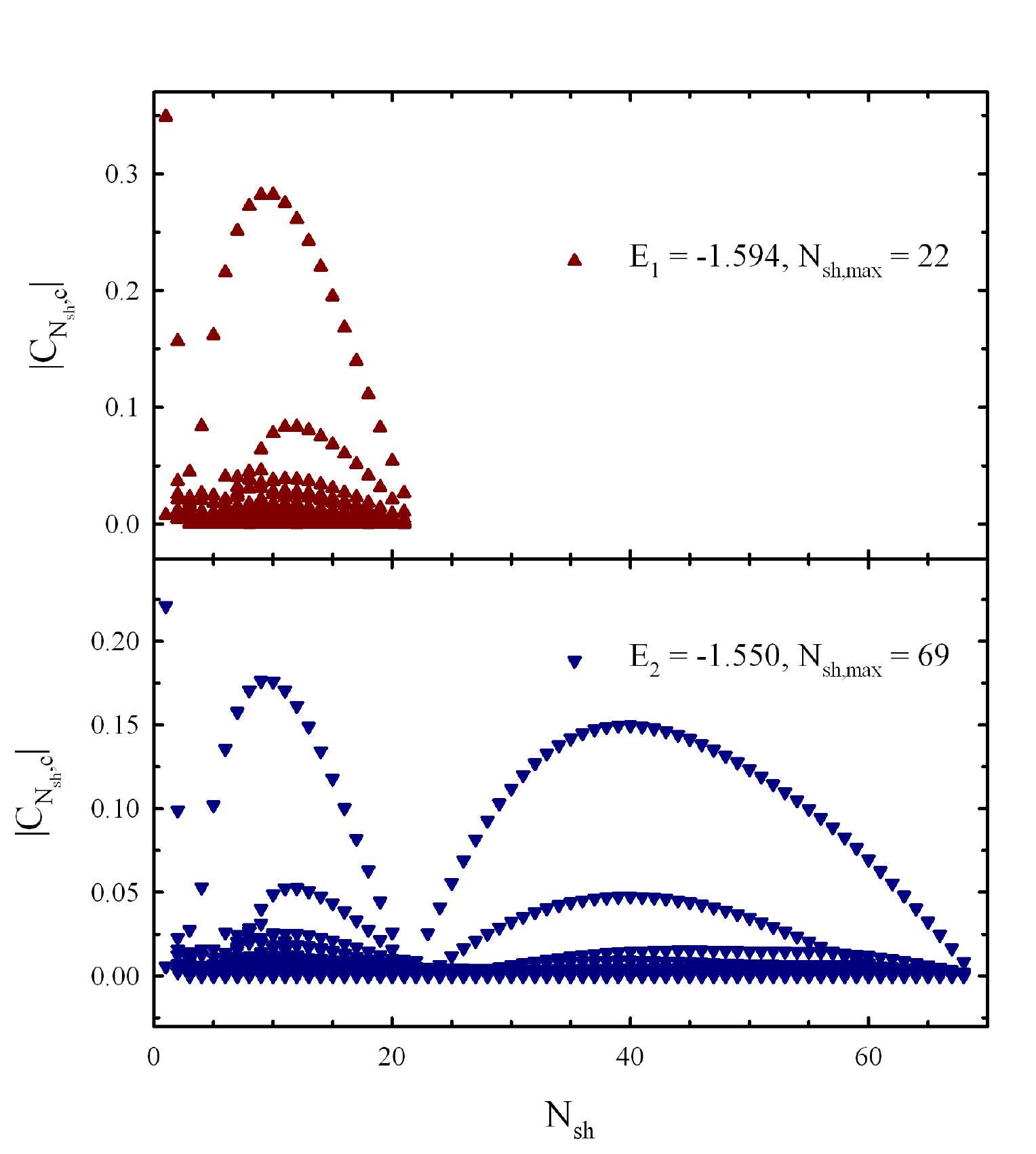}%
\caption{Comparison of wave functions  obtained with the different number of oscillator functions 
but approximately with the same energy.}
\label{Fig:WaveFuns7LiComp}%
\end{center}
\end{figure}

Fig. \ref{Fig:WaveFuns7LiComp} demonstrates that wave functions with the same energy but obtained
with the different number of oscillator functions are similar to each other in the range of smaller
value of $N_{sh}$ and differ only by normalization factor. These wave functions represent
3/2$^{-}$ states of $^{7}$Li and are calculated with $K_{\max }$=13.
Note that 982
oscillator functions participate in the construction of the first wave function
(upper panel), while 3614 oscillator functions are involved in the expansion of
the second wave function (lower panel). It is noteworthy, that when
we use oscillator functions up to $N_{sh,\max }$, it allows us to describe
and to analyze wave functions and intercluster distance in the range $0\leq
\rho \leq b\sqrt{4N_{sh,\max }+2L+6}$. This inequality reflects very
important feature of oscillator functions. As easy to see, the second wave
function repeats the behaviour of the first wave function in the range $%
0\leq N_{sh}\leq 22$, and presents new part of the wave function for larger
values of $N_{sh}$: $22<N_{sh}\leq 69$.

In Table \ref{Tab:Distan7Be32M} we present average distances between clusters
in $^{7}$Li ground and excited 3/2$^{-}$ states, calculated with $K_{\max}%
$=13. These quantities determine the most probable shape of a triangle joining
the centers of mass of interacting clusters. How to calculate the average
distances is explained in Refs. \cite{2013UkrJPh.58.544V,2014PAN..77.555N}.
For each tree, $R_{2}$ stands for the average distance between a pair of
clusters indicated in brackets, and $R_{1}$ determines mean distance between
the first cluster and the center of mass of the two-cluster subsystem. One can
see that the ground state is a compact state in all threes of the Jacobi
coordinates. The first excited state, as  expected, has dominant
$^{4}$He+$^{3}$H structure, since $^{3}$H nuclei as a binary subsystem $^{3}%
$H=$n+d$ is very compact and is located far away (8.13 fm) from $^{4}$He.
Quite similar structure is observed for the second excited state. This state,
as one can see in Fig. \ref{Fig:Spectr7Li32MK7_13}, lies below the $n+^{6}%
$Li threshold. In this state, the size of $^{3}$H\ is slightly increased to
4.19 fm compared to the ground and the first excited states, and distance
between $^{3}$H and $^{4}$He exceeds 9 fm. The third excited state is of
different nature. It is located between two-cluster ($n+^{6}$Li) and the
three-cluster thresholds, and thus has very distinguished two-cluster $n+^{6}%
$Li structure. Indeed, in this state the spacing between clusters d and $^{4}%
$He, comprising $^{6}$Li, is 4.42 fm, slightly more than in the ground state
($\approx$3 fm); and the distance from neutron to the center of mass $^{6}$Li
is very large - more than 17 fm. It is important to recall that the first,
second and other excited states belong to two- or three-cluster
continua. The wave functions of these states, as was pointed out earlier, are
normalized to unity within the selected size of a "discrete \ box". Thus,
Table \ref{Tab:Distan7Be32M} presents not absolute values of intercluster
distances, but their relative values.%

\begin{table}[tbp] \centering
\begin{ruledtabular}
\caption{Average distances between clusters for the ground and excited $3/2^-$
states in $^7$Li.}%
\begin{tabular}
[c]{ccccc}
Tree & \multicolumn{2}{c}{$\alpha+\left(  n+d\right)  $} &
\multicolumn{2}{c}{$n+\left(  d+\alpha\right)  $}\\\hline
$E$, MeV & $R_{1}$ & $R_{2}$ & $R_{1}$ & $R_{2}$\\\hline
-8.688 & 2.88 & 2.46 & 2.70 & 2.99\\
$E_{th}\left(  ^{3}H+\alpha\right)  $=-6.654 &  &  &  & \\
-3.939 & 8.13 & 3.12 & 5.81 & 8.03\\
-1.644 & 9.68 & 4.19 & 6.69 & 9.19\\
$E_{th}\left(  n+^{6}Li\right)  $=-1.475 &  &  &  & \\
-0.439 & 6.85 & 18.06 & 17.38 & 4.42\\
1.263 & 8.80 & 10.20 & 16.05 & 4.71\\
\end{tabular}
\label{Tab:Distan7Be32M}%
\end{ruledtabular}
\end{table}%

\begin{figure}[ptbh]
\begin{center}
\includegraphics[width=\columnwidth]%
{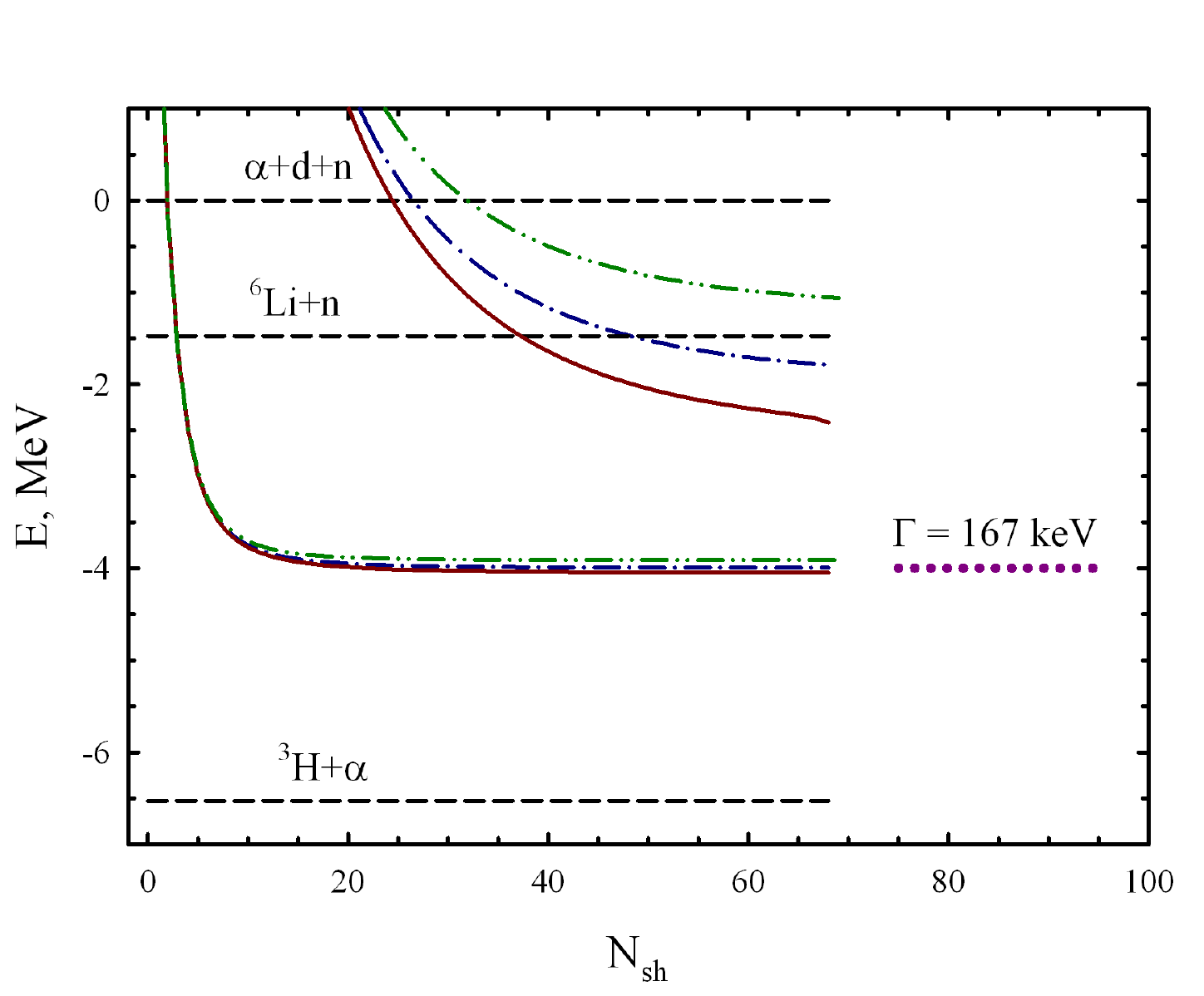}%
\caption{Spectrum of the 7/2$^{-}$ states in $^{7}$Li as a function of
$N_{sh}$ calculated with $K_{\max}$ =9 (dash-dot-dot lines), $K_{\max}$=11
(dash-dot lines) and$K_{\max}$=13 (solid lines). Dot line indicates a position
of the 7/2$^{-}$ resonance state in $^{7}$Li calculated with an alternative
microscopical method (see text for details).}%
\label{Fig:Spectr7Li72MK9_13}%
\end{center}
\end{figure}

In Figure \ref{Fig:Spectr7Li72MK9_13} we display trajectories of eigenenergies
of $^{7}$Li for the 7/2$^{-}$ state. These trajectories are calculated with
$K_{\max}$=9, $K_{\max}$=11  and
$K_{\max}$=13. One can see that the lowest eigenstates have a
plateau at energy $E$= -3.91 MeV ($K_{\max}$=11), $E$= -3.99 MeV ($K_{\max}%
$=11) and $E$= -4.05 MeV ($K_{\max}$=13). These energies are practically
stable when we change $N_{sh}$ from 45 to 70. Such a plateau may indicate that
there is a narrow resonance states at this energy. This is a very simple but
reliable way for locating a position of a narrow resonance state which is used
in the Stabilization Method \cite{1970PhRvA...1.1109H}. To be sure, we made
use of an alternative method to calculate the energy and width of the
resonance state in two-body continuum. We refer to this method as AM GOB
(Algebraic Model for scattering which involves Gaussian and Oscillator Basis
to describe relative motion of three clusters), it imposes proper boundary
conditions for scattering of the third cluster on a bound state of two-cluster
subsystem. This method was formulated in Ref. \cite{2009NuPhA.824...37V} and
applied to study bound and resonance states in $^{7}$Be. Investigation of
discrete and continuous spectrum in $^{7}$Li within the AM GOB method was
carried out in Ref. \cite{2009PAN....72.1450N}. We adopted this model to study
resonance states in $^{7}$Li with the same input parameters which are used in
the AM HHB model. Results of these calculations are shown in Figure
\ref{Fig:Spectr7Li72MK9_13} by a dashed line indicating energy of the 7/2$^{-}$
resonance state. We observe a very good agreement between these two methods.
Thus our method (AM HHB), even with a rather restricted basis of
hyperspherical functions, correctly predicts position of the 7/2$^{-}$
resonance state. It is worthwhile underlying that the stabilization method
works perfectly only for the narrow resonance states. This the case for the
7/2$^{-}$ resonance state as its total width is equal to 167 keV.

To understand the essence of the method used and to determine the impact of
$K_{\max}$ and $N_{sh}$ on the obtained results, we consider the geometry of
the three-cluster system $\alpha+d+n$ in the 7/2$^{-}$ states of $^{7}$Li. As
it was demonstrated, three approximations $K_{\max}$=9 , $K_{\max}$=11 and
$K_{\max}$=13 give a very close energies of the lowest 7/2$^{-}$ states. In
these cases one can see explicitly and unambiguously the role $K_{\max}$ and
$N_{sh}$ in description of \ three-cluster nuclei with the hyperspherical
harmonics. For this aim, in Table \ref{Tab:7LiRadii72M} we show the energy and
average distances between clusters for the 7/2$^{-}$ states of $^{7}$Li calculated with $K_{\max}%
$=9, $K_{\max}$=11 and $K_{\max}$=13 and $N_{sh}=70$. \ We use the same
notations as in Table \ref{Tab:Distan7Be32M}, however here we restricted ourselves with 
 $\alpha+d+n$ tree, which represents the lowest two-body channel $\alpha+t$.
Note that difference of the plateau energies, calculated with $K_{\max}$=9 and
$K_{\max}$=13 is 138 keV which is less than the width ($\Gamma$=167 keV) of
the 7/2$^{-}$ resonance state. By increasing the number of channels from
$K_{\max}$=9 to $K_{\max}$=13, we slightly improve the description of the
$t=d+n$ subsystem as the average distance $R_{2}$ between a deuteron and a neutron is
somewhat increased. When we increase $N_{sh}$ from minimal to maximal value,
we allow a three-cluster system to have more larger distance between all
clusters or between the third cluster and a compact (i.e. bound) state of
two-cluster subsystem.%

\begin{table}[tbp] \centering
\begin{ruledtabular}
\caption{The energy and  average distances between clusters in the 7/2$^-$ states 
of $^7$Li calculated with $K_{\max}$ =9, $K_{\max}$ =11, and $K_{\max}$ =13.}%
\begin{tabular}
[c]{cccc}
$K_{\max}$ & $E$, MeV & $R_{1}$, fm & $R_{2}$, fm\\\hline
9 & -3.911 & 2.813 & 2.657\\
11 & -3.993 & 2.976 & 2.664\\
13 & -4.049 & 3.224 & 2.684\\
\end{tabular}
\label{Tab:7LiRadii72M}%
\end{ruledtabular}
\end{table}%

From the presented Figures and Tables, and from the analysis of wave functions of
the states in two-body continuum we conjecture that the more channels of the
hyperspherical harmonics (or larger values of $K_{\max}$) are involved, the
more precise is the description of two-cluster bound states. By increasing
$N_{sh}$ with a fixed value of $K_{\max}$, we allow the third cluster to move
far away from the bound two-cluster subsystem.

The conclusions, we made by analysing Tables \ref{Tab:Distan7Be32M} and \ref{Tab:7LiRadii72M}, can be
confirmed by considering wave functions of the ground and excited states in
the coordinate or oscillator representations.

Let us consider the correlation functions, \ which we define as%
\[
D_{E_{\nu}}\left(  x_{\alpha},y_{\alpha}\right)  =\sum_{\lambda_{\alpha
},l_{\alpha}}\left\vert \psi_{E_{\nu};\lambda_{\alpha},l_{\alpha};L}^{\left(
\alpha\right)  }\left(  x_{\alpha},y_{\alpha}\right)  \right\vert^2 \ x_{\alpha
}^{2}y_{\alpha}^{2}.%
\]
For an adequate physical interpretation of the correlation function, we
introduce $s_{\alpha},r_{\alpha}$ distances between clusters instead of the
Jacobi coordinates $x_{\alpha},y_{\alpha}$%
\[
\mathbf{s}_{\alpha}=\sqrt{\frac{A_{\beta}+A_{\gamma}}{A_{\beta}A_{\gamma}}%
}\mathbf{x}_{\alpha},\quad\mathbf{r}_{\alpha}=\sqrt{\frac{A_{\alpha}+A_{\beta
}+A_{\gamma}}{A_{\alpha}\left(  A_{\beta}+A_{\gamma}\right)  }}\mathbf{y}%
_{\alpha}. %
\]
We will show correlation functions $D_{E_{\nu}}\left(  s_{\alpha},r_{\alpha
}\right)  $ for bound and pseudo-bound states. \ We will also display sections of
the correlation functions $D_{E_{\nu}}\left(  s_{\alpha},r_{\alpha,\max
}\right)  $ and $D_{E_{\nu}}\left(  s_{\alpha,\max},r_{\alpha}\right)  $,
where ($s_{\alpha,\max},r_{\alpha,\max}$) is a point in two-dimensional plane
($s_{\alpha},r_{\alpha}$) determining a principal maximum of the correlation
function for an eigenstate of three-cluster hamiltonian with energy $E_{\nu}$.
Coordinates ($s_{\alpha},r_{\alpha}$) of the principal and local maxima of the correlation
function determine maximal probability to find a three-cluster system at this point, thus we will call them as dominant distances between clusters and  we will refer to the quantity $s_{\alpha}$ as to a dominant size of a two-cluster subsystem.

\begin{figure}[ptbh]
\begin{center}
\includegraphics[width=\columnwidth]%
{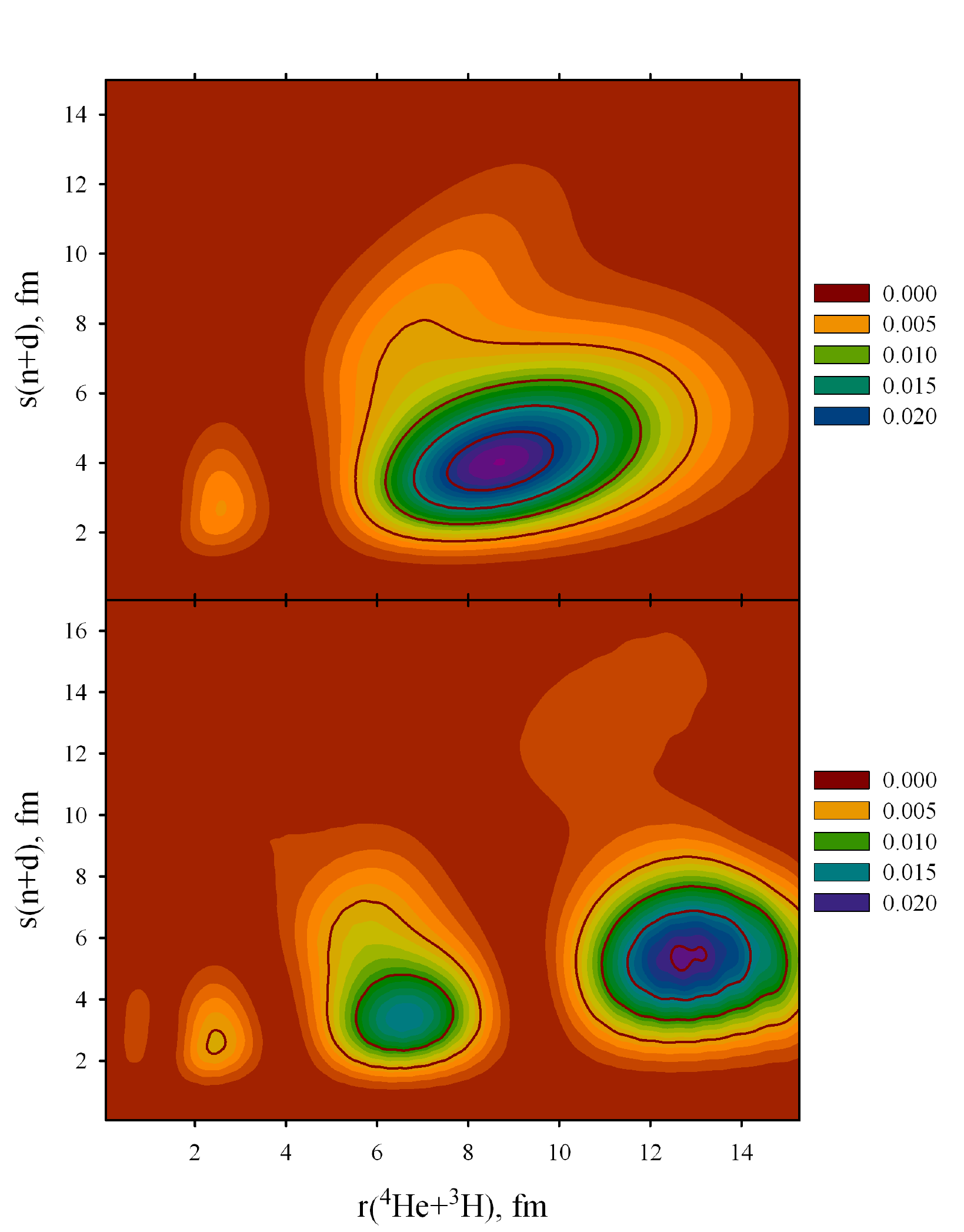}%
\caption{The correlations functions for the first excited state (upper part)
and for the second excited state (lower part) in $^{7}$Li.}%
\label{Fig:CorrFuns7Li32ME1E2}%
\end{center}
\end{figure}

The contour plots of the correlation functions for the first and second
excited states in $^{7}$Li are presented in Fig.
\ref{Fig:CorrFuns7Li32ME1E2}. This figure is constructed for the first Jacobi
tree, where $\mathbf{s}_{1}$\ determines distance between neutron and deuteron
while vector $\mathbf{r}_{1}$ represents distance between an alpha particle
and $^{3}$H nucleus. It is important to recall that both states belong to the
$^{4}$He+$^{3}$H continuum. One may notice that the larger is the energy of the
excited state, the more oscillations of the correlation functions are observed
along the distance $r_{1}$ within the presented range. One may also see that
the larger is the distance between an alpha particle and $^{3}$H, the larger is
the dominant size of $^{3}$H or, in other words, the larger is the distance
between deuteron and neutron comprising nucleus $^{3}$H. Such behaviour of the
correlation functions indicates the polarizability of $^{3}$H as a two-cluster
subsystem when the distance between $^{3}$H and alpha particle is changed.
These results are in \ accordance with the results of Refs.
\cite{2009NuPhA.824...37V, 2009PAN....72.1450N}, where the cluster
polarizability of two-cluster subsystems in $^{7}$Be and $^{7}$Li were
investigated with the model designed to study effects of cluster polarization
on the structure of bound and resonance states in compound systems.%

\begin{figure}[ptbh]
\begin{center}
\includegraphics[width=\columnwidth]%
{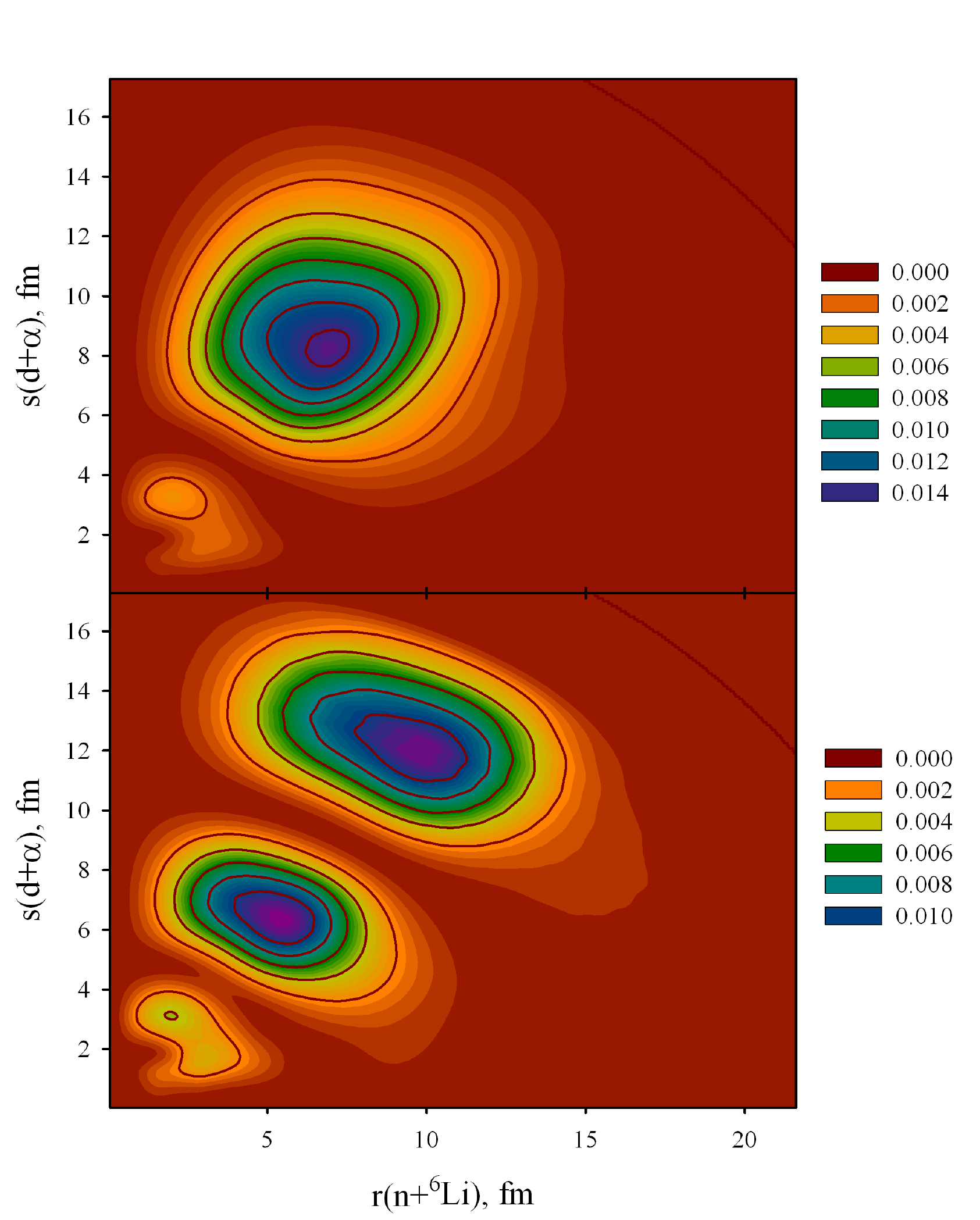}%
\caption{The correlation functions for the first and second excited states in
$^{7}$Li obtained for the second Jacobi tree.}%
\label{Fig:CorrFuns7Li32MT2E1E2}%
\end{center}
\end{figure}

In the second Jacobi tree with the dominant two-cluster configuration $n+^{6}%
$Li, we obtain rather different view (Fig. \ref{Fig:CorrFuns7Li32MT2E1E2})
of the correlations functions for the same excited states in $^{7}$Li. In this
representation, the size of $^{6}$Li and distance between neutron and $^{6}$Li
nucleus are very large. This is in a compliance with Fig.
\ref{Fig:CorrFuns7Li32ME1E2}, where the dominant distance between deuteron and
neutron is small, however the dominant distances between neutron and alpha
particle and between deuteron and alpha particle are very large.%

\begin{figure}[ptbh]
\begin{center}
\includegraphics[width=\columnwidth]%
{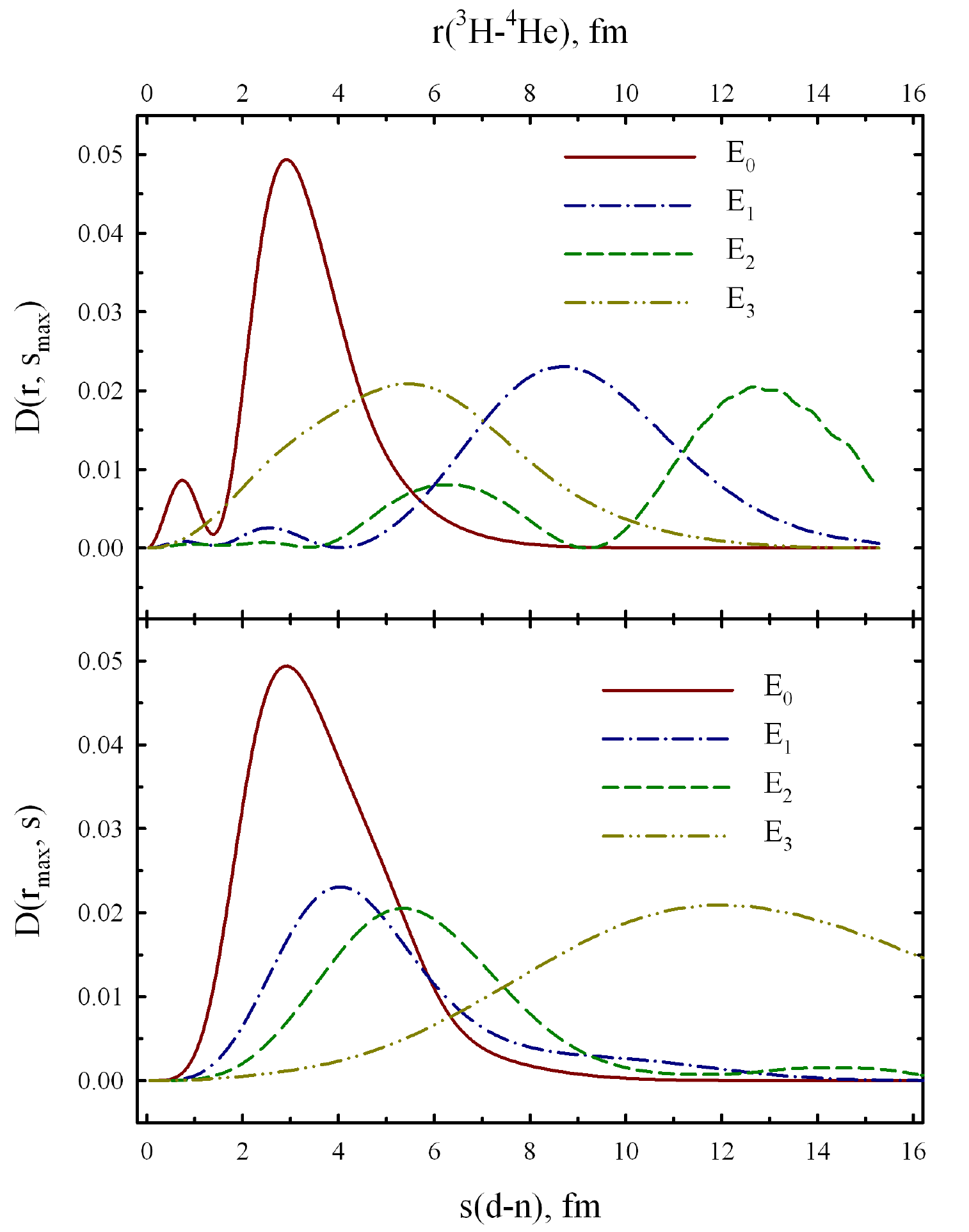}%
\caption{Parts of the correlation functions for the $J^{\pi}$=3/2$^{-}$ ground
and excited states of $^{7}$Li.}%
\label{Fig:WaveFuns7Li32M}%
\end{center}
\end{figure}

In Fig. \ref{Fig:WaveFuns7Li32M} we display sections $D_{E_{\nu}}\left(
s_{\alpha},r_{\alpha,\max}\right)  $ and $D_{E_{\nu}}\left(  s_{\alpha,\max
},r_{\alpha}\right)  $ of the correlation functions for the lowest states of
$^{7}$Li with the total angular momentum $J^{\pi}$=3/2$^{-}$. \ The lower part
of Fig. \ref{Fig:WaveFuns7Li32M} shows that nucleus $^{3}$H, comprised of a
deuteron and a neutron, is confined in states $E_{0}$, $E_{1}$ and $E_{2}$.
With increasing energy, the distance
between deuteron and neutron is increased from 3 to 5.5 fm. As in Fig.
\ref{Fig:CorrFuns7Li32ME1E2}, such behaviour of wave functions of $^{3}$H in
$E_{0}$, $E_{1}$ and $E_{2}$ states more explicitly shows a polarization of
$^{3}$H cluster when it interacts with an alpha-particle. The upper part of
Fig. \ref{Fig:WaveFuns7Li32M} \ demonstrates that $^{7}$Li is a compact
object in its ground state, since the dominant distance between clusters
$^{3}$H and $^{4}$He is approximately 3 fm. Wave functions and correlation
functions of the excited states $E_{1}$ and $E_{2}$ have an oscillatory
behaviour along coordinate $r_{1}$, determining distance between clusters
$^{3}$H and $^{4}$He. As one can see in Fig. \ref{Fig:Spectr7Li32MK7_13},
these states belong to the $^{3}$H+$^{4}$He continuous spectrum.

Fig. \ref{Fig:WaveFuns7Li32M} completely confirms the conclusions which were
made by analysing the results presented in Table \ref{Tab:Distan7Be32M}.%

The behaviour of the presented correlations functions is totally in agreement
with the asymptotic forms of three-cluster wave functions, discussed above in
Eqs. (\ref{eq:011})-(\ref{eq:015A}). Indeed, for the bound state we observe the
exponential tails of the correlation functions along the vectors
$\mathbf{s}_{\alpha}$ and $\mathbf{r}_{\alpha}$, while for the states in the
two-body continuum the exponential tail is observed along the vector
$\mathbf{s}_{\alpha}$ and the oscillating tail is seen along the vector
$\mathbf{r}_{\alpha}$. These conclusions are correct not only for $^{7}$Li but
also for all other nuclei considered in this paper.

Concluding this section we note that we have carried out similar
investigations for a mirror nucleus $^{7}$Be as three-cluster configuration
$^{7}$Be=$\alpha+d+p$. The Coulomb interaction, which is more stronger in $^{7}$Be, 
slightly changes energy of the two-cluster threshold $^{4}$He+$^{3}$He and reduces 
the energy of the $^{7}$Be ground state with respect to two- and three-cluster thresholds.
Therefore, all results and
conclusions deduced for $^{7}$Li nucleus are valued for the mirror nucleus.
For the lack of room in the present paper, we will not dwell on the results for
$^{7}$Be.

\subsection{$^{8}$Be=$^{4}$He+$^{3}$H+$p$}

\begin{figure}[ptbh]
\begin{center}
\includegraphics[width=\columnwidth]%
{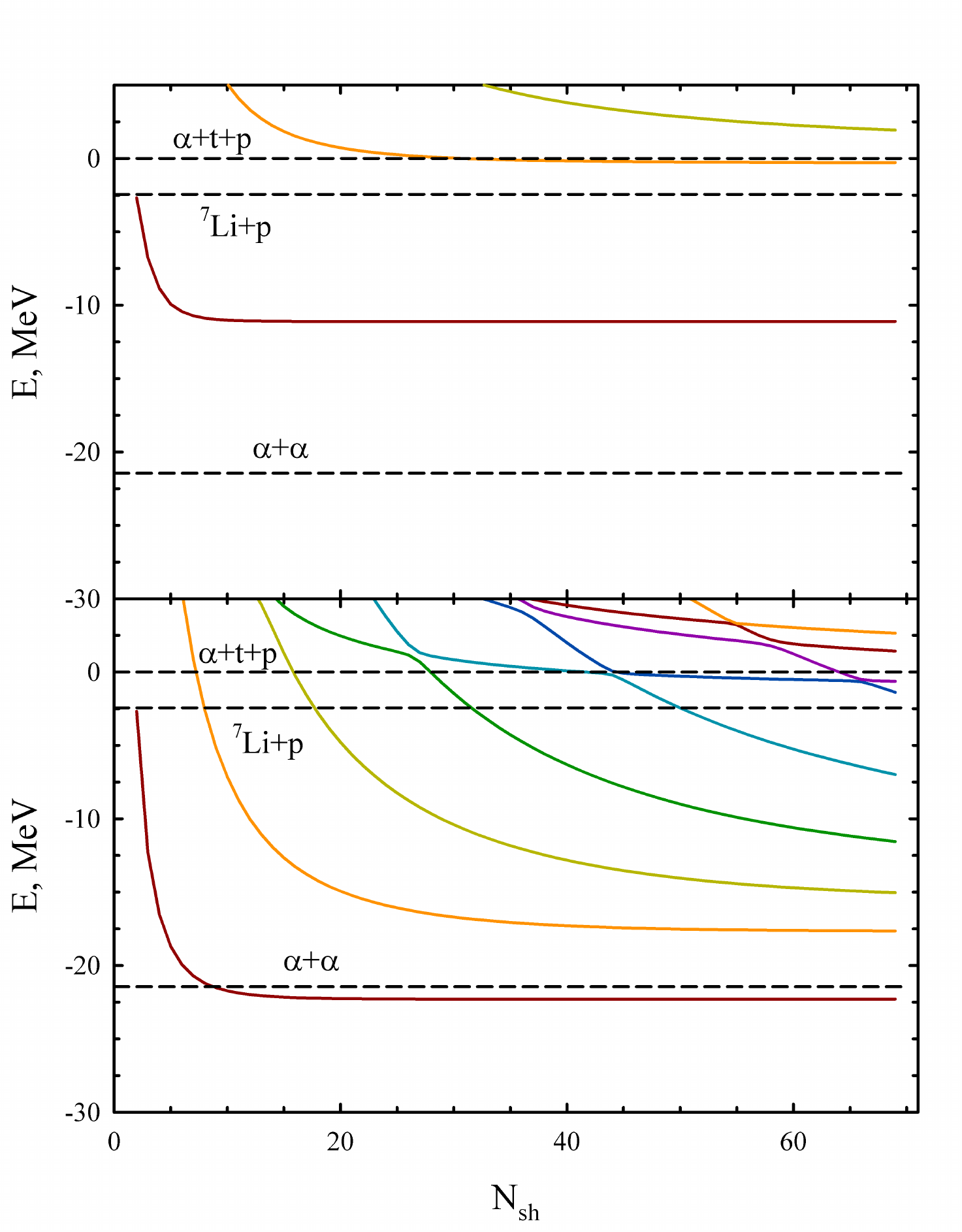}%
\caption{Spectrum of $0^{+}$ states in $^{8}$Be as a function of $N_{sh}$,
calculated with $K_{\max}=$0 (upper part) and $K_{\max}=$14 (lower part of the
figure).}%
\label{Fig:Spectr8Be0PK0K14}%
\end{center}
\end{figure}

Now we consider spectrum and wave functions of the $0^{+}$ state in $^{8}$Be.
With the three-cluster configuration $^{4}$He+$^{3}$H+$p$ we have got the
following binary channels: $^{4}$He+$^{4}$He and $^{7}$Li+$p$. We do not
consider the binary channel $^{5}$Li+$^{3}$H as its threshold energy exceeds
the three-cluster threshold. The energy of $0^{+}$ states in $^{8}$Be,
calculated with only one hyperspherical harmonic $K=0$ and with $K_{\max}=$14,
as a function of $N_{sh}$ is displayed in Fig. \ref{Fig:Spectr8Be0PK0K14}.
The first and important result is that only one hyperspherical harmonic
($K$=0) is able to produce one state in the $^{4}$He+$^{4}$He continuum and
one state above the $^{7}$Li+$p$ threshold but below the three-cluster $^{4}%
$He+$^{3}$H+$p$ threshold. Besides, the
"ground" $0^{+}$ state appears in the two-cluster $^{4}$He+$^{4}$He continuum
starting with $N_{sh}=$2, while the first
excited state needs more than $N_{sh}=$30 oscillator shells to appear in the the $^{7}$Li+$p$ continuum.
It is interesting to note (see the lower part of
Fig. \ref{Fig:Spectr8Be0PK0K14}), that hyperspherical harmonics with
$K_{\max}=$14 generate one bound state (below the $^{4}$He+$^{4}$He threshold)
and four states in two-cluster $^{4}$He+$^{4}$He continuum and also two states
in the $^{7}$Li+$p$ continuum. Thus this number of hyperspherical harmonics
(i.e. all harmonics with 0$\leq K\leq$14 or 36 channels) is able to describe
some states of elastic $^{4}$He+$^{4}$He and $^{7}$Li+$p$ scattering and the
reaction $^{4}$He+$^{4}$He $\Longleftrightarrow$ $^{7}$Li+$p$ at two discrete
energy points. It should be stressed that these results are obtained without
imposing the boundary conditions.

Average distances between clusters in $^{8}$Be are shown in Table
\ref{Tab:Distan8Be0P}. They are calculated with $K_{\max}=$14.We can see that
the binary channel $^{4}$He+$^{4}$He dominates in pseudo-bound states within
energy range from -17.648 to -1.396 \ MeV (energy of these states are measured from
the three-cluster threshold). In all these states, cluster $^{4}$He comprised
of triton and proton is very compact. The last state, listed in Table
\ref{Tab:Distan8Be0P}, has the dominant $^{7}$Li+$p$ structure. Note that the
energy of the $^{7}$Li+$p$ threshold is -2.45 MeV. Thus states with energies
$E$=-1.396 MeV and $E$=-0.634 MeV lie above the $^{7}$Li+$p$ threshold, the
first of them has $^{4}$He+$^{4}$He as a dominant structure, while the second
one is mainly represented by the $^{7}$Li+$p$ structure.%

\begin{table}[htbp] \centering
\begin{ruledtabular}
\caption{Average distances between clusters for  $0^+$
states in $^8$Be.}%
\begin{tabular}
[c]{ccccc}
Tree & \multicolumn{2}{c}{$\alpha+\left(t+p\right)$} &
\multicolumn{2}{c}{$p+\left(t+\alpha\right)$}\\\hline
$E$, MeV & $R_{1}$ & $R_{2}$ & $R_{1}$ & $R_{2}$\\\hline
-22.298 & 2.168 & 2.49 & 2.67 & 2.13\\
-17.648 & 4.71 & 3.07 & 3.98 & 4.68\\
-15.042 & 5.89 & 3.72 & 4.89 & 5.84\\
-11.561 & 5.97 & 3.81 & 4.96 & 5.94\\
-6.993 & 5.80 & 2.93 & 4.24 & 5.82\\
$E_{th}\left(  p+^{7}Li \right)  $=-2.134 &  &  &  & \\
-1.396 & 5.66 & 2.82 & 4.21 & 5.70\\
-0.634 & 5.13 & 15.38 & 15.19 & 4.00\\
\end{tabular}
\label{Tab:Distan8Be0P}%
\end{ruledtabular}
\end{table}%

\subsection{$^{4}$He=$d+p+n$}

\begin{figure}[ptbh]
\begin{center}
\includegraphics[width=\columnwidth]%
{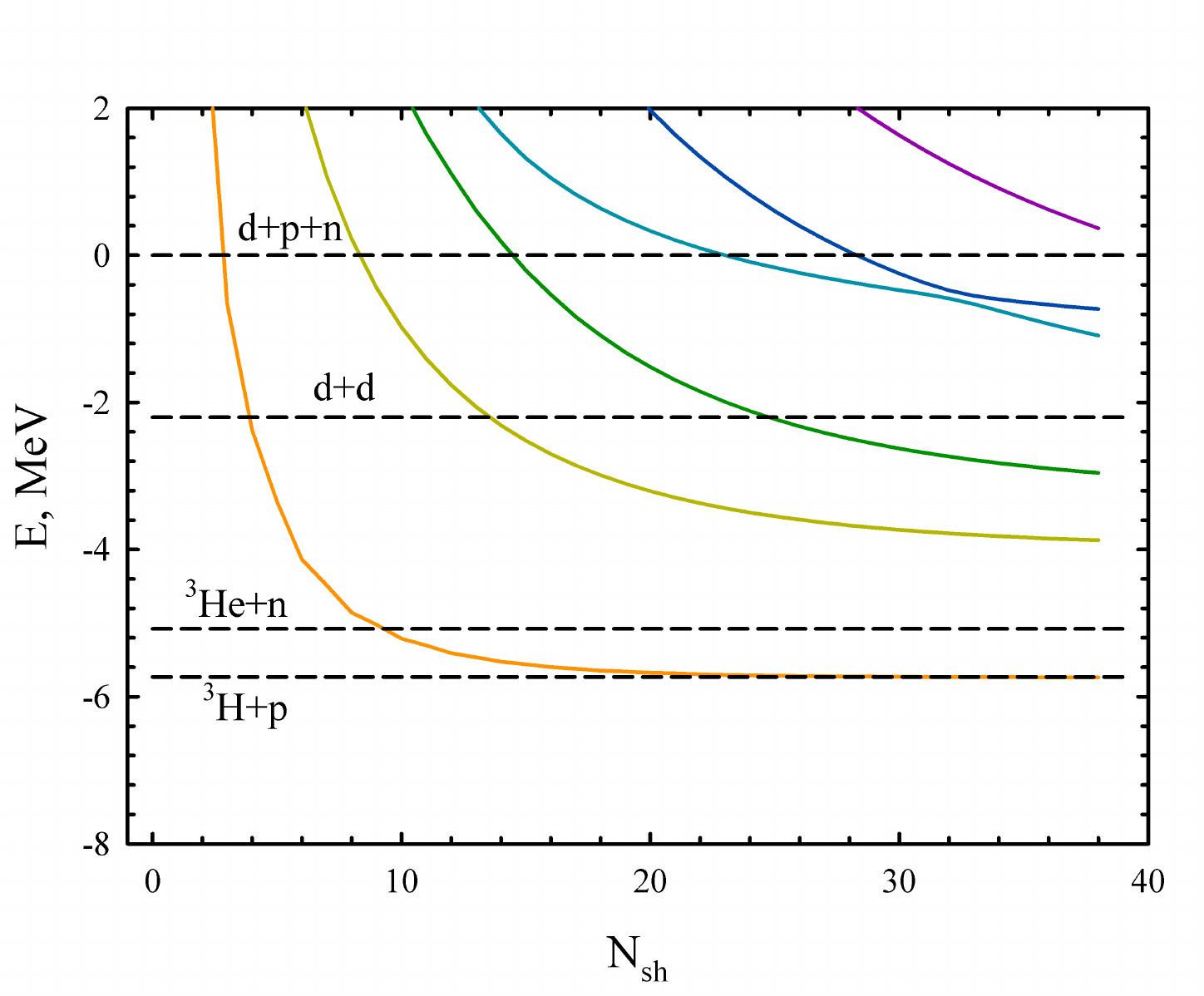}%
\caption{Spectrum of the $0^{+}$ state in $^{4}$He calculated with $K_{\max
}=14$.}%
\label{Fig:Spectr4HeK14}%
\end{center}
\end{figure}

We consider $^{4}$He as a three-cluster configuration $^{4}$He=$d+p+n$. This
configuration allows us to take into account all binary channels of $^{4}$He,
namely, $^{3}$H+$p$, $^{3}$He+$n$ and $d+d$. In Table \ref{Tab:Distan4He0P} we
show spectrum of $^{4}$He and average distance between clusters, obtained with
$K_{\max}=$14. The first state is the ground state in $^{4}$He, and it is very
compact structure being a deeply bound state. All excited states belong to
different two-cluster continua. Positions of these states with respect to
two-body thresholds are shown in Fig. \ref{Fig:Spectr4HeK14}. One can see in
Table \ref{Tab:Distan4He0P} that all excited states are very dispersed.%

\begin{table}[htbp] \centering
\begin{ruledtabular}
\caption{Average distances between clusters for  $0^+$
states in $^4$He}%
\begin{tabular}
[c]{ccccccc}
Tree & \multicolumn{2}{c}{$p+\left(  d+n\right)  $} &
\multicolumn{2}{c}{$n+\left(  d+p\right)  $} &
\multicolumn{2}{c}{$d+\left(  n+p\right)  $}\\\hline
$E$, MeV & $R_{1}$ & $R_{2}$ & $R_{1}$ & $R_{2}$ & $R_{1}$ & $R_{2}$\\\hline
-28.089 & 1.08 & 1.26 & 1.07 & 1.27 & 1.13 & 1.15\\
$E_{th}\left(  d+n\right)  =$-5.732 &  &  &  &  &  & \\
-5.738 & 5.13 & 4.42 & 4.42 & 5.21 & 3.74 & 6.13\\
$E_{th}\left(  d+p\right)  =$-5.077 &  &  &  &  &  & \\
-3.873 & 8.40 & 6.36 & 6.78 & 8.25 & 5.35 & 10.12\\
-2.958 & 7.72 & 8.79 & 9.00 & 7.29 & 5.85 & 11.13\\
$E_{th}\left(  n+p\right)  =$-2.202 &  &  &  &  &  & \\
-1.090 & 8.99 & 7.83 & 8.19 & 8.76 & 6.05 & 11.40\\
-0.732 & 6.56 & 9.24 & 6.85 & 9.00 & 8.79 & 4.89\\
\end{tabular}
\label{Tab:Distan4He0P}%
\end{ruledtabular}
\end{table}%

\subsection{$^{10}$Be}

Let us consider spectrum of $^{10}$Be, provided that $^{10}$Be is treated as a  
$\alpha +\alpha +^{2}n$ three-cluster configuration, and analyze what is the most 
probable geometry of this three-cluster
structure. In Table \ref{Tab:spectr10Be} we show the the ground and the first 
excited $0^{+}$ states in $^{10}$Be. The results
are obtained with the MP. In Ref. \cite{Lashko201778} the exchange parameter
$u$ of the potential was selected so to reproduce the energy of the $^{10}$Be
ground state with respect to the binary threshold $^{6}$He$+\alpha .$ We use
the same value of this parameter. With this value of the parameter $u$ we
obtained the relative position of the threshold energies of the binary $^{6}$He$
+\alpha $ and \ $^{8}$Be$+^{2}n$ channels indicated in Table \ref{Tab:spectr10Be}.

\begin{table}[htbp] \centering
\begin{ruledtabular}
\caption{Energies of the ground and the lowest excited $0^+$
states in $^{10}$Be. Dominant two-body channels and their threshold energies $E_{th}$ are also presented.}
\begin{tabular}
[c]{cc}
2Cluster threshold & $E$, MeV\\\hline
& -9.162\\
& -2.197\\
$^{6}$He+$\alpha$ & $E_{th}=$-1.745\\
& -0.343\\
$^{8}$Be+$^{2}n$ & $E_{th}=$-0.023\\
& 0.561\\
& 1.162\\
\end{tabular}
\label{Tab:spectr10Be}
\end{ruledtabular}
\end{table}

As can be seen from Table \ref{Tab:spectr10Be}, the ground state and the 
first excited $0^+$ state are below the lowest binary decay threshold of $^{10}$Be. 
The second excited state lies between $^{6}$He+$\alpha$ and $^{8}$Be+$^{2}n$ thresholds, 
while the rest of the states belong to three-cluster continuum.

\begin{figure}[ptbh]
\begin{center}
\includegraphics[width=\columnwidth]{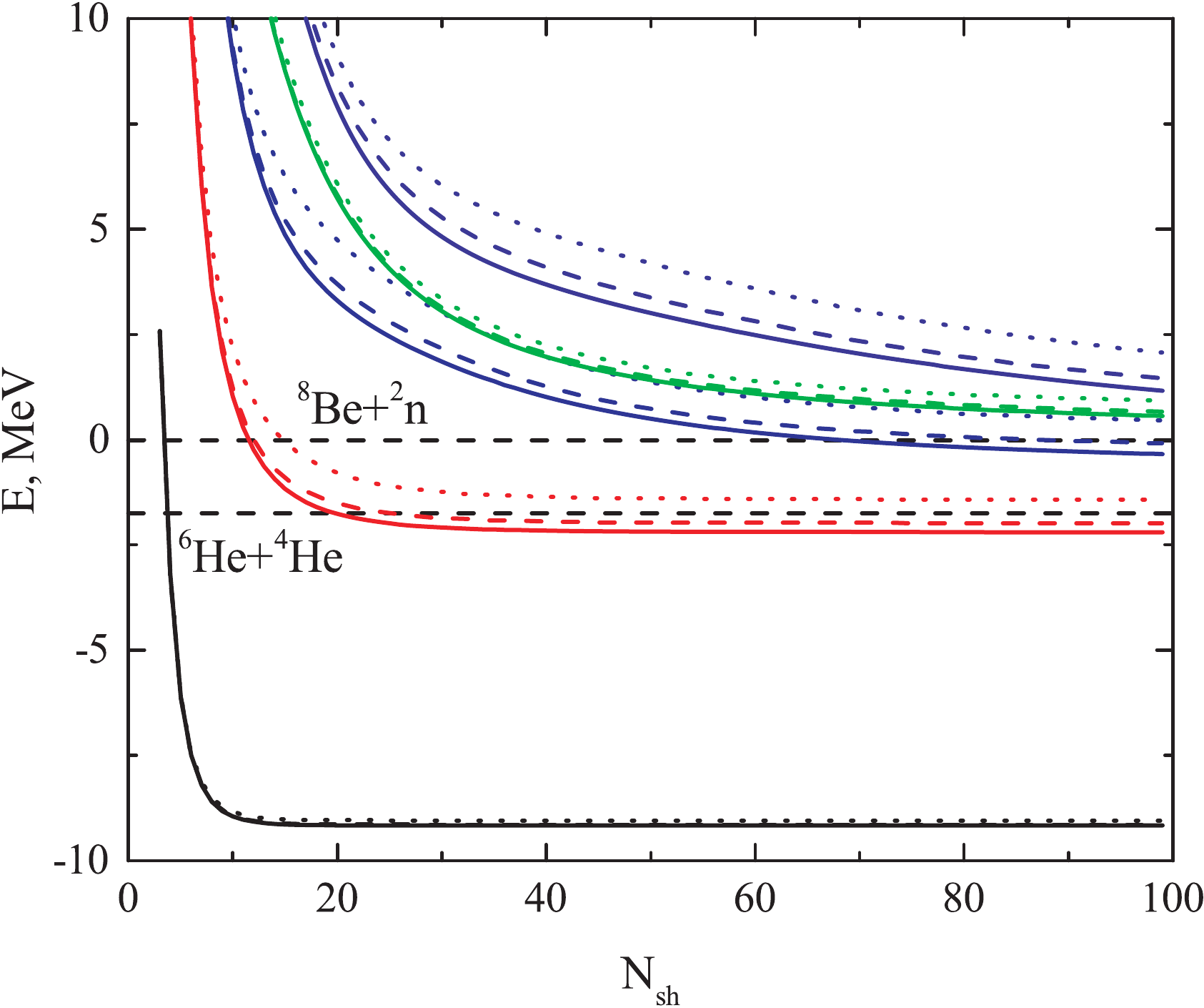}
\end{center}
\caption{Spectrum of the 0$^{+}$
states in $^{10}$Be as a function of $N_{sh}$ \ and $K_{\max }$. Dotted lines correspond to $K_{\max }=6$, dashed lines denote $K_{\max }=10$, and solid lines stand for $K_{\max }=14$.}
\label{Fig:spectrum10Be}
\end{figure}

Spectrum of the 0$^{+}$
states in $^{10}$Be as a function of the number of oscillator shells and maximum 
value of hypermomentum involved in the calculations  is plotted in Fig. \ref{Fig:spectrum10Be}.
 As evident from Fig. \ref{Fig:spectrum10Be}, to reproduce the energy of the ground state 
it is sufficient to invoke basis functions with $N_{sh}=20$ and $K_{\max }=6$. The higher 
is the energy of the state, the larger value of the number of oscillator shells and hypermomentum 
should be used to reach the convergence. However, the third excited state with energy $E=0.56$ MeV 
above the three-cluster decay threshold of $^{10}$Be somewhat differs from the other states presented 
in Fig. \ref{Fig:spectrum10Be}. Hyperharmonics with $K_{\max }\geq10$ slightly contribute to the 
energy of this state as opposed to the neighbouring excited states.

\begin{figure}[ptbh]
\begin{center}
\includegraphics[width=\columnwidth]{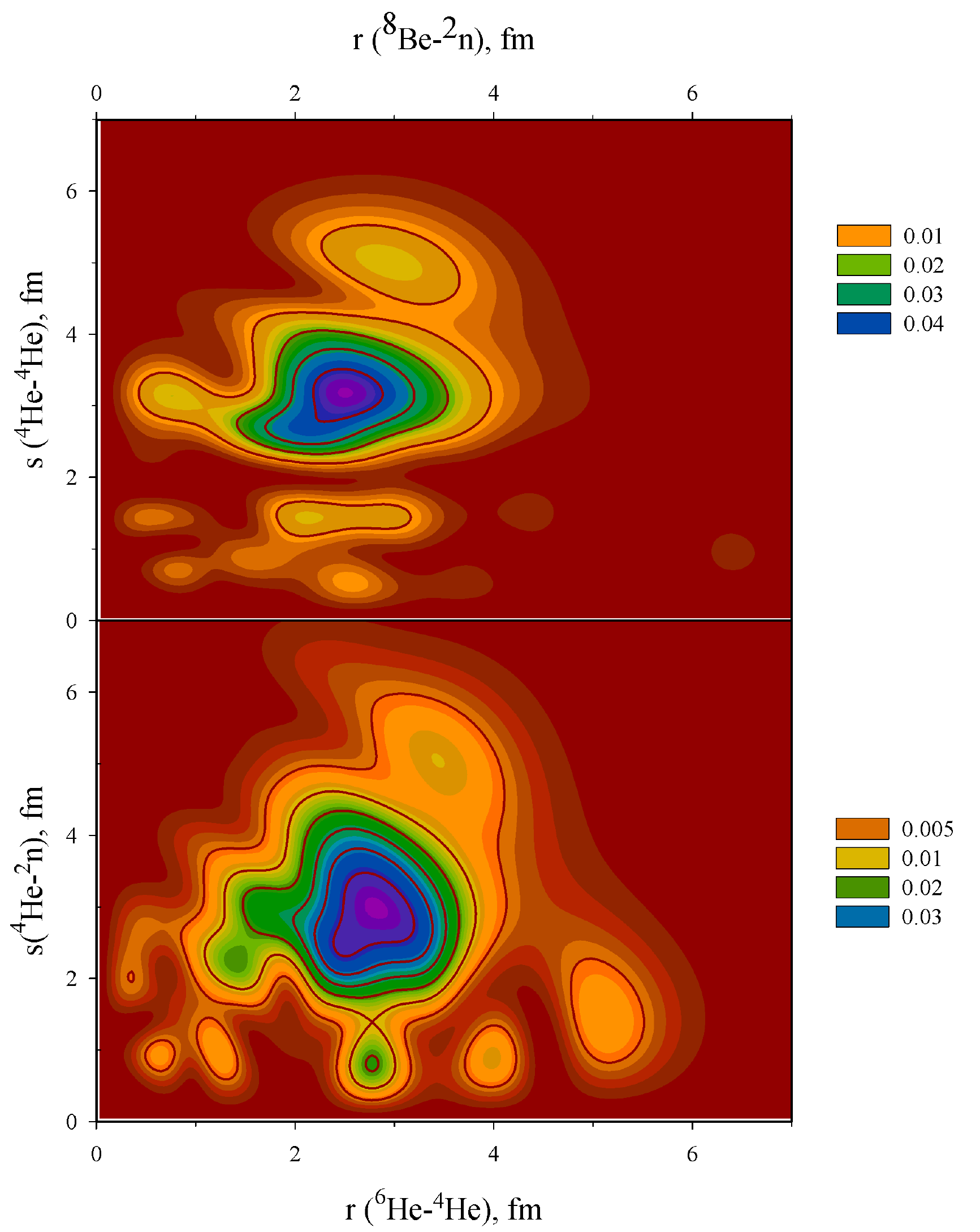}
\end{center}
\caption{The correlation functions for the ground state in
$^{10}$Be in different Jacobi trees.}
\label{Fig:gs10Be}
\end{figure}

\begin{figure}[ptbh]
\begin{center}
\includegraphics[width=\columnwidth]{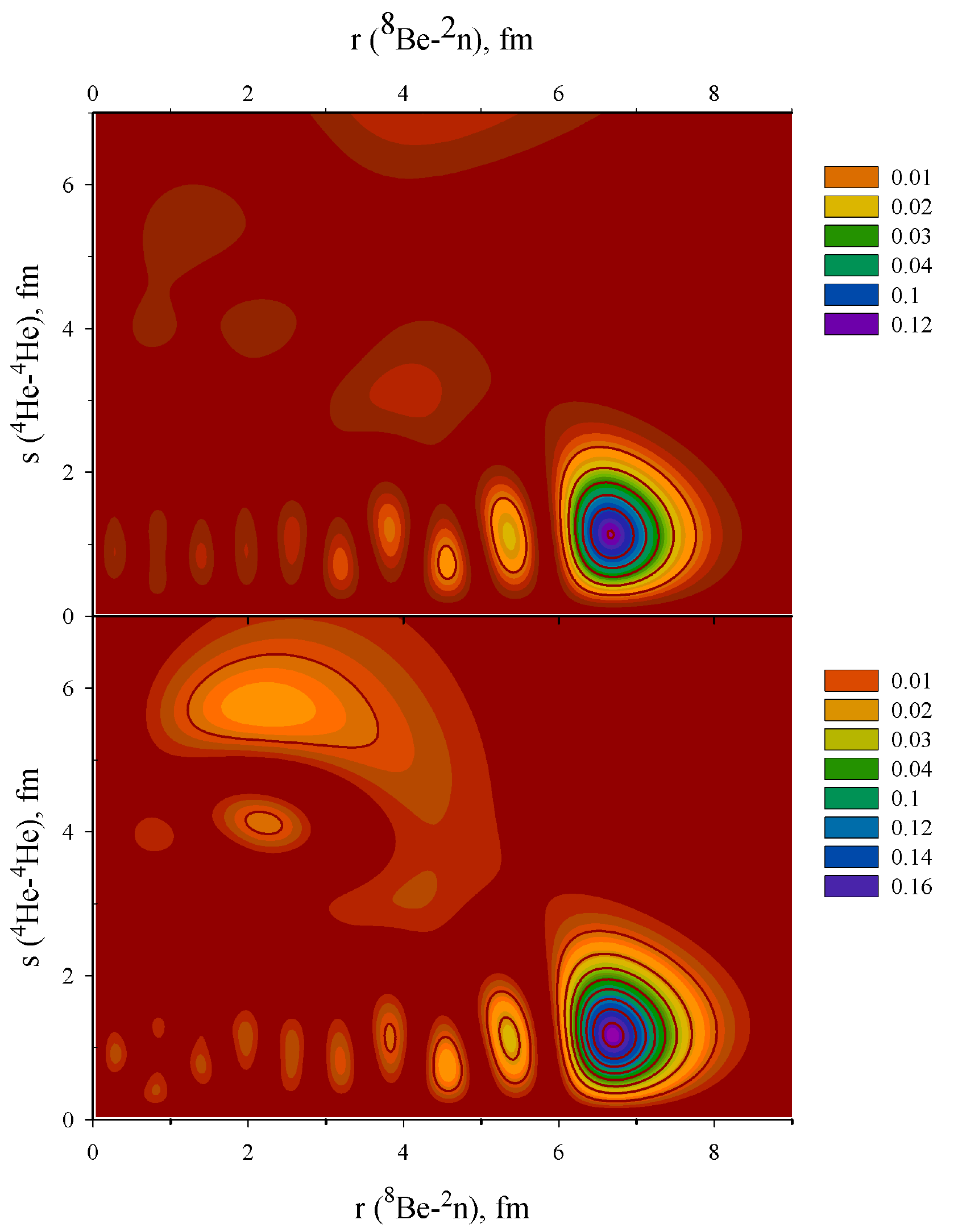}
\end{center}
\caption{The correlation functions for the first excited state (lower part)
and for the second excited state (upper part) in $^{10}$Be.
}
\label{Fig:E12_10Be}
\end{figure}

\begin{figure}[ptbh]
\begin{center}
\includegraphics[width=\columnwidth]{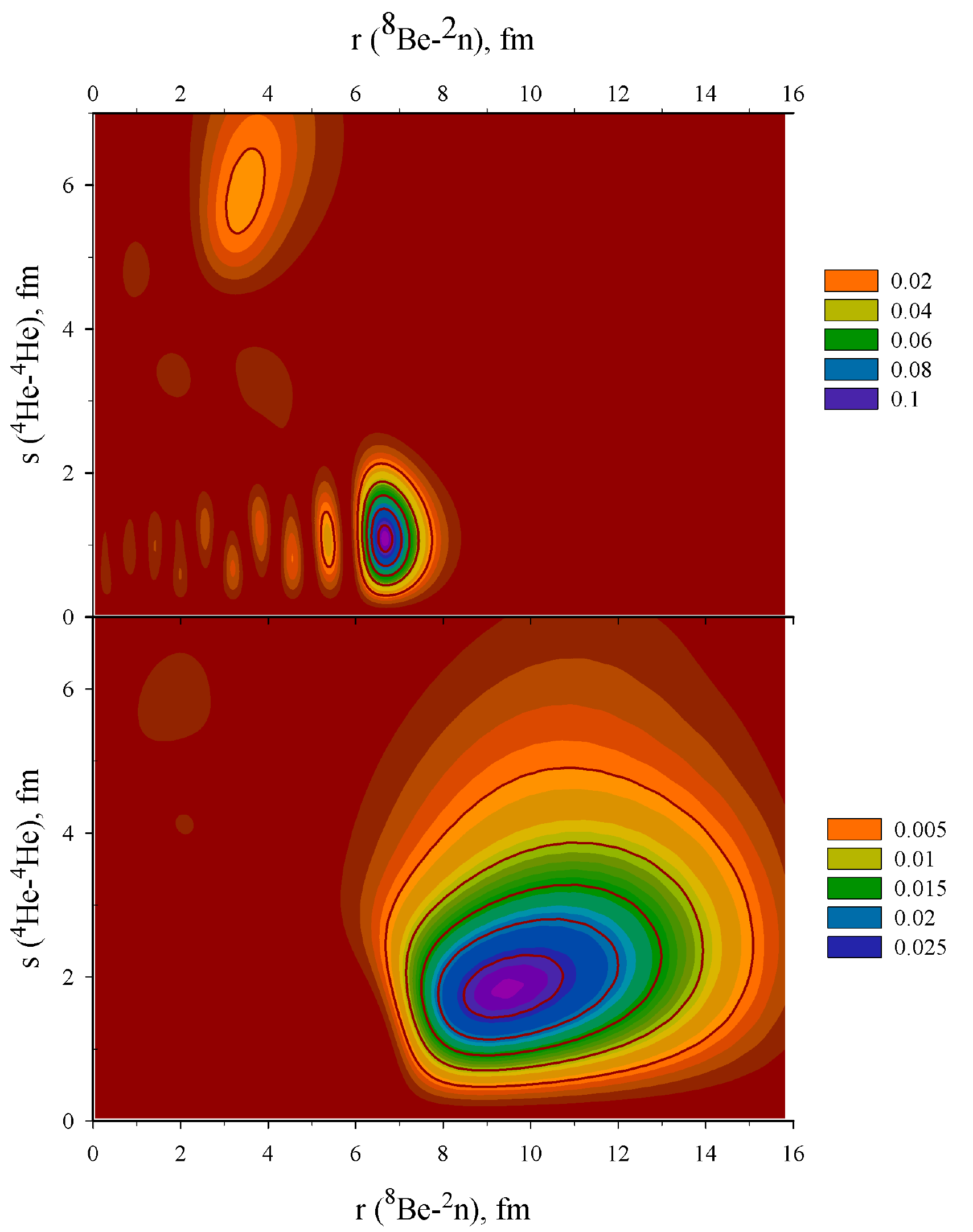}
\end{center}
\caption{The correlation functions for the third excited state (lower part)
and for the fourth excited state (upper part) in $^{10}$Be.
}
\label{Fig:E34_10Be}
\end{figure}

Figs. \ref{Fig:gs10Be}, \ref{Fig:E12_10Be}, \ref{Fig:E34_10Be} display the contour plots of 
the correlation functions for the ground and excited states in $^{10}$Be tabulated in Table 
\ref{Tab:spectr10Be}.
Referring to Fig. \ref{Fig:gs10Be}, we can conclude that $^{10}$Be in its ground state is 
a system of three equally spaced clusters (two $\alpha$-particles and a dineutron). The other 
four excited states, contrastingly, have prominent $^{8}$Be$+^2$n structure, as is clear 
from Figs. \ref{Fig:E12_10Be} and \ref{Fig:E34_10Be}. $^{6}$He$+\alpha$ configuration 
reveals itself only at higher energies.

\begin{figure}[ptbh]
\begin{center}
\includegraphics[width=0.8\columnwidth]{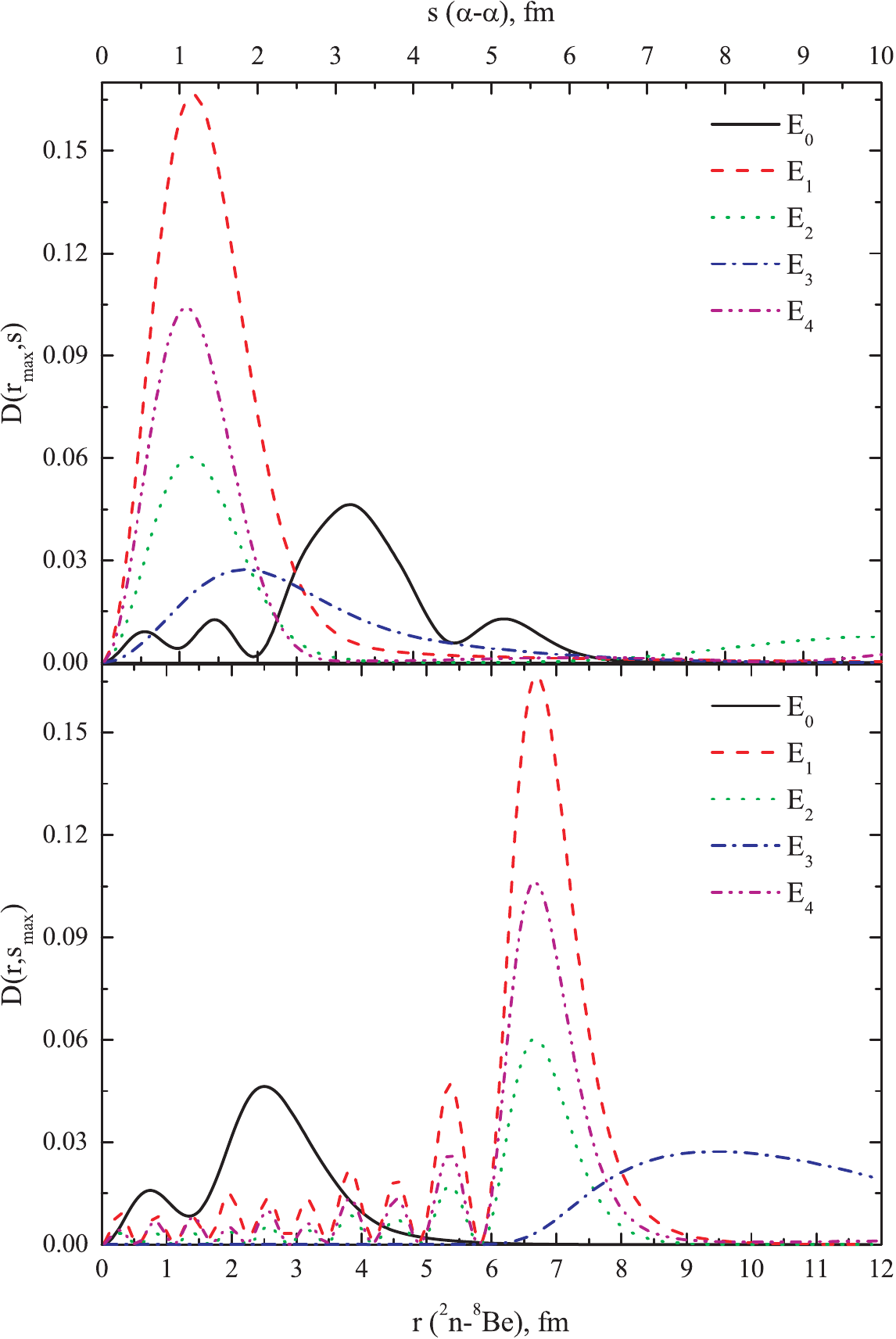}
\end{center}
\caption{Parts of the correlation functions for the $J^{\pi}$=0$^{+}$ ground
and excited states of $^{10}$Be.}
\label{Fig:corrfun10Be}
\end{figure}

Sections of the correlation functions, presented in Figs. \ref{Fig:gs10Be}, \ref{Fig:E12_10Be}, 
\ref{Fig:E34_10Be}, are shown in Fig. \ref{Fig:corrfun10Be}. As may be inferred from Fig. 
\ref{Fig:corrfun10Be}, in three low-lying states of the $^{10}$Be the $^{8}$Be subsystem is 
rather compact as compared to the distance between $^{8}$Be and a dineutron. As for the first 
state which is above the three-cluster decay threshold, it is characterized by somewhat more 
dilute $^{8}$Be subsystem, but $^{8}$Be$+^2$n configuration still dominates in this state.

\begin{figure}[ptbh]
\begin{center}
\includegraphics[width=0.8\columnwidth]{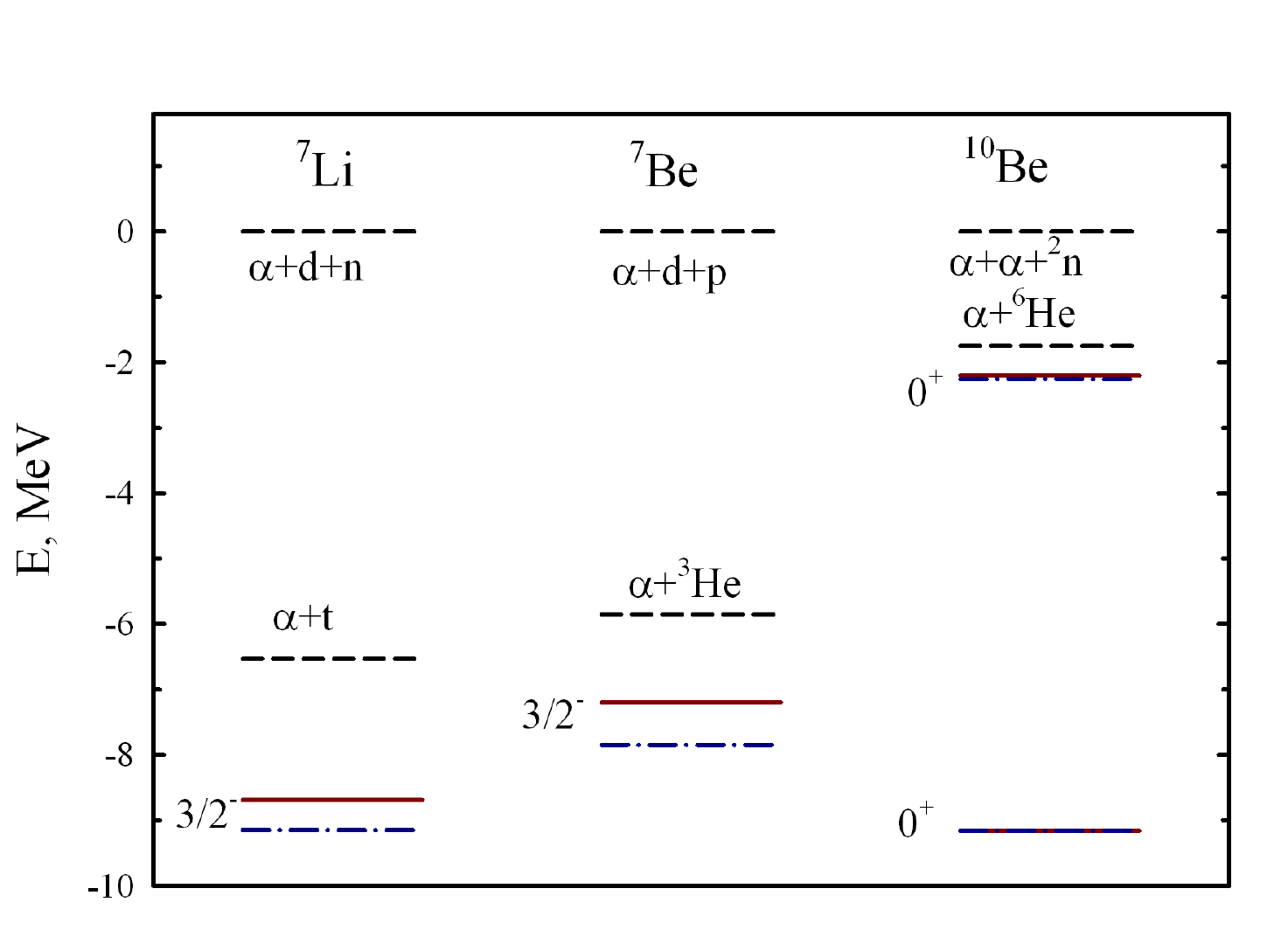}
\end{center}
\caption{Energies of the 3/2$^-$ ground states in $^7$Li and $^7$Be, and the ground and first 
excited 0$^+$ states in $^{10}$Be calculated within the present model (solid line) and within 
the AM GOB model (dash-dotted line).}
\label{Fig:SpectrComp}
\end{figure}

We did not consider in detail the convergence which the present model provides for the bound 
states in the selected nuclei. However, some conclusions can be made from Fig. \ref{Fig:Spectr7Li32MK7_13}, 
where we display the spectrum of the 3/2$^-$ states in $^7$Li calculated with the different values of 
$K_{\max}$ and $N_{sh}$. To demonstrate that the present model provides us with a satisfactory 
description of the bound states, in Fig. \ref{Fig:SpectrComp} we show the ground state energy of 
$^7$Li and $^7$Be and the energies of the ground and first excited 0$^+$ states in $^{10}$Be. We 
compare results of the present model with the corresponding results of the AM GOB model mentioned above. 
The AM GOB model involves a little larger part of the total Hilbert space to describe the discrete and 
continuous spectrum of a three-cluster system, which is reduced to a set of the binary channels. 
In Refs. \cite{2009PAN....72.1450N}, \cite{2009NuPhA.824...37V} and \cite{Lashko201778} the AM GOB 
model have been applied to study nuclei $^7$Li,  $^7$Be  and $^{10}$Be  with the same input parameters, 
which we use in the present model. Therefore, we will consider results of the model as ``exact''. 
In Fig. \ref{Fig:SpectrComp} we also display by dashed lines both the three-cluster and the lowest 
two-cluster thresholds. We can see that all ground states are deeply bound states with respect to 
the three-cluster threshold, however their convergence strongly depends on how far these states lie 
with respect to the lowest two-cluster channel. The latter, as we pointed above, determines the shape 
of the asymptotic part of the bound state wave function. As one could expect, the Coulomb interaction 
reduces the bound state energy  in $^7$Be (comparing to $^7$Li) and slows down the convergence of 
this state. By concluding, the presented results show that our  model is able to reproduce the bound 
state energy of the considered nuclei with a satisfactory precision.

\section{Conclusion \label{Sec:Conclusions}}

Within a microscopic three-cluster model we have considered spectra of a set
of light nuclei: $^{4}$He, $^{7}$Li, $^{7}$Be, $^{8}$Be, $^{10}$Be. We
selected those nuclei which have a dominant three-cluster channel and one or
more two-body channels below the three-cluster decay threshold. We considered tree 
kinds of nuclei. Two of these nuclei are deeply bound ($^{4}$He and $^{10}$Be), as 
their ground states lie below -8 MeV with respect to the lowest two-cluster threshold. 
Two other nuclei ($^{7}$Li and $^{7}$Be) are weekly bound, since their binding energy 
does not exceed -2.6 MeV. The last nucleus  $^{8}$Be has no bound states. A full set of
the antisymmetric three-cluster oscillator functions was constructed. These
functions were labelled by the quantum numbers of the hyperspherical harmonics
method. Matrix elements of a hamiltonian, consisting of central
nucleon-nucleon forces and the Coulomb potential, between the oscillator
functions were constructed and eigenvalues and corresponding eigenfunctions
were calculated. We analysed dependence of the eigenvalues on the number of
oscillator functions involved in calculations. It was demonstrated that some
of the eigenvalues are the discrete states in two-cluster continuum. Analysis
of the wave functions in coordinate and oscillator representations showed that
these functions have a correct asymptotic behaviour peculiar to the two-cluster
continuous spectrum.

The main result of the present investigations is that it is possible to see the 
evidence of two-cluster structure in the three-cluster wave function of a pseudo-bound 
state yet with 
a rather restricted set of the hyperspherical
harmonics and hyperradial excitations as well. It was demonstrated that the
 eigenstates of the three-cluster hamiltonian have correct
asymptotic behaviour both for bound states below two-cluster threshold and
states in two-cluster continuum. Analysis of the correlation functions in
different Jacobi trees reveals polarizability of two-cluster bound states when
the distance between the third cluster and two-cluster subsystem is relatively small.

\section*{Acknowledgment}

This work was supported in part by the Program of Fundamental
Research of the Physics and Astronomy Department of the National
Academy of Sciences of Ukraine (Project No. 0117U000239) and by the Ministry of Education and Science
of the Republic of Kazakhstan, Research Grant No. IPS 3106/GF4.

\end{document}